\documentclass[journal=inoraj,manuscript=article]{achemso}

\usepackage{amsmath}
\usepackage{graphicx}
\usepackage{bm}

\newcommand{\ket}[1]{\ensuremath{\left|#1\right\rangle}}
\newcommand{\bra}[1]{\ensuremath{\langle#1|}}

\title{Nature of hyperfine interactions in TbPc$_2$ single-molecule magnets: Multireference \emph{ab-initio} study}

\author{Aleksander L. Wysocki}
\affiliation{Department of Physics, Virginia Tech, Blacksburg, Virginia 24061, United States}
\email{alexwysocki2@gmail.com}

\author{Kyungwha Park}
\affiliation{Department of Physics, Virginia Tech, Blacksburg, Virginia 24061, United States}
\email{kyungwha@vt.edu}

\begin{document}

\begin{abstract}
Lanthanide-based single-ion magnetic molecules can have large magnetic hyperfine interactions as well as large magnetic anisotropy. Recent experimental studies reported tunability of these properties by changes of chemical environments or by application of external stimuli for device applications. In order to provide insight onto the origin and mechanism of such tunability, here we investigate the magnetic hyperfine and nuclear quadrupole interactions for $^{159}$Tb nucleus in TbPc$_2$ (Pc=phthalocyanine) single-molecule magnets using multireference {\it{ab-initio}} methods including spin-orbit interaction. Since the electronic ground and first-excited (quasi)doublets are well separated in energy, the microscopic Hamiltonian can be mapped onto an effective Hamiltonian with an electronic pseudo-spin $S=1/2$. From the {\it ab-initio}-calculated parameters, we find that the magnetic hyperfine coupling is dominated by the interaction of the Tb nuclear spin with electronic orbital angular momentum. The asymmetric $4f$-like electronic charge distribution leads to a strong nuclear quadrupole interaction with significant non-axial terms for the molecule with low symmetry. The {\it ab-initio} calculated electronic-nuclear spectrum including the magnetic hyperfine and quadrupole interactions is in excellent agreement with experiment. We further find that the non-axial quadrupole interactions significantly influence the avoided level crossings in magnetization dynamics and that the molecular distortions affect mostly the Fermi contact terms as well as the non-axial quadrupole interactions.
\end{abstract}

\AtEndDocument{
\newpage
\begingroup
\section*{TOC Graphic}
\sffamily
\singlespacing
\begin{center}
\fbox{
\begin{minipage}{3.25in}
\vbox to 1.75in{\includegraphics{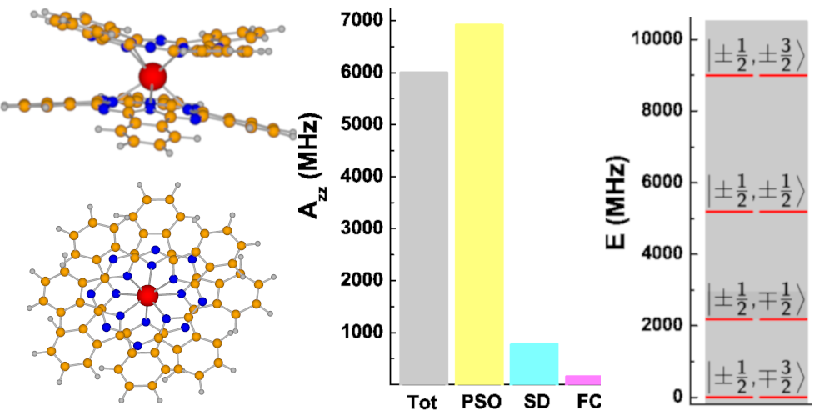}}
\end{minipage}
}
\end{center}
\endgroup
}

\section{Introduction}

Lanthanide-based single-ion magnetic molecules \cite{Sessoli2009,Baldovi2012,Woodruff2013,Liddle2015,Wang2016,Guo2018,Goodwin2017,Thiele2014,Godfrin2017}
have drawn a lot of attention due to superior magnetic properties, structural compatibility with substrates, and proof-of-concept experiments
for magnetic information storage and quantum information science applications. Recently, molecular crystals of dysprosium-based single-ion
magnetic molecules have shown magnetic hysteresis above liquid nitrogen temperature \cite{Guo2018} and an experimental effective energy barrier over 1000~cm$^{-1}$ \cite{Goodwin2017}. Especially, double-decker [LnPc$_2$] (Pc=phthalocyanine) single-molecule magnets (SMMs) \cite{Ishikawa2003} are shown to form periodic layers on substrates \cite{YangHe2014,Wackerlin2016,Studniarek2019}. Furthermore, the terbium (Tb) nuclear magnetic moment
was found to be strongly coupled to the electronic degrees of freedom in TbPc$_2$ SMMs \cite{Ishikawa2004,Vincent2012,Urdampilleta2013,Thiele2013}.
By taking advantage of a strong hyperfine Stark effect, the Tb nuclear levels in a TbPc$_2$ SMM have been used for experimental realizations of
Rabi oscillations \cite{Thiele2014}, Ramsey interferometry \cite{Godfrin2018}, and Grover's algorithm \cite{Godfrin2017} within a single-molecule transistor platform.


\begin{figure}[t!]
\centering
\includegraphics[width=1.0\linewidth]{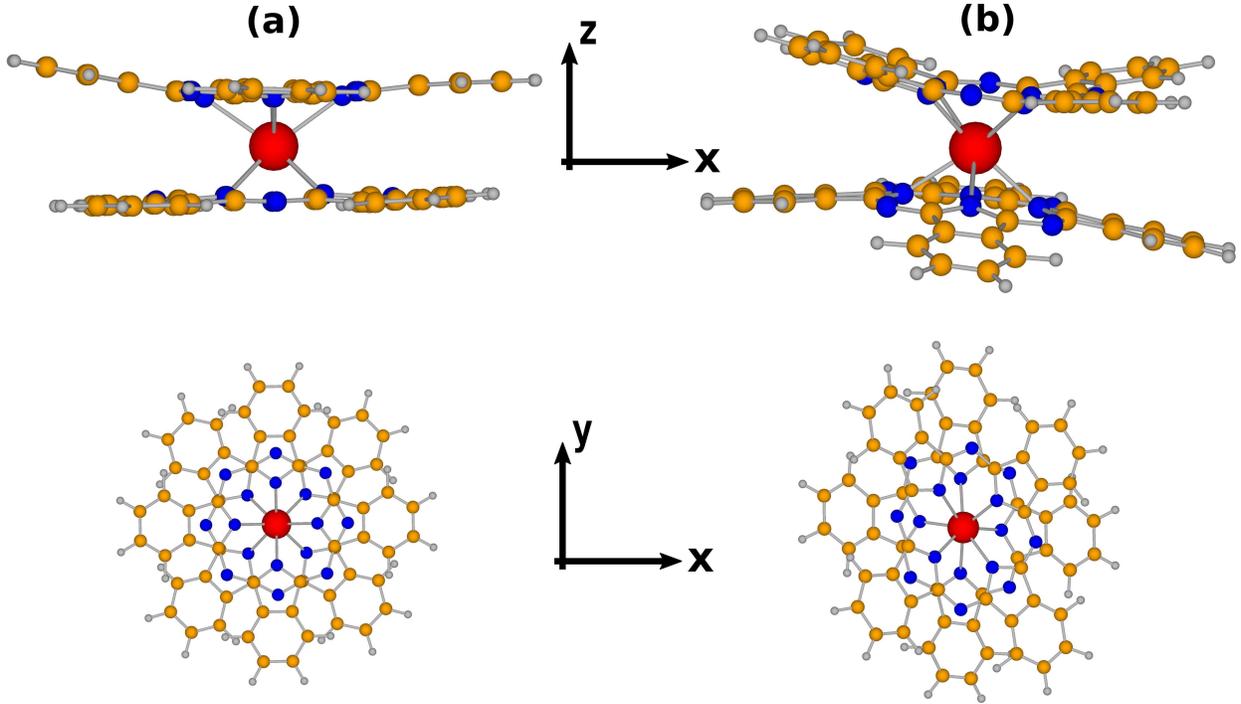}
\caption{Side and top views of experimental structure of two TbPc$_2$ molecules of interest. (a) Neutral TbPc$_2$ with geometry from Ref.~\citenum{Komijani2018}. (b) Anionic TbPc$_2$ with geometry from Ref.~\citenum{Branzoli2009}. Red, blue, orange, and gray spheres represent
Tb, N, C, and H atoms, respectively. The coordinate system corresponds to magnetic axes obtained by diagonalization of the $\mathbf{g}$ matrix calculated for the ground (quasi)doublet for each molecule.}
\label{Geometry}
\end{figure}

In TbPc$_2$ SMMs, one Tb$^{3+}$ ion is sandwiched between two approximately planar Pc ligands (Fig.~\ref{Geometry}). Three different oxidation states ($-1$, 0, and $+1$) were experimentally realized \cite{Ishikawa2004b,Loosli2006,Takamatsu2007,Branzoli2009,Katoh2009,Ganivet2013,Komijani2018}.
The synthesized TbPc$_2$ molecules have only approximate $D_{4d}$ symmetry. The degree of symmetry deviation varies with crystal packing, diamagnetic dilution molecules, or solvent molecules used in synthesis processes.



\begin{figure}[t!]
\centering
\includegraphics[width=1.0\linewidth]{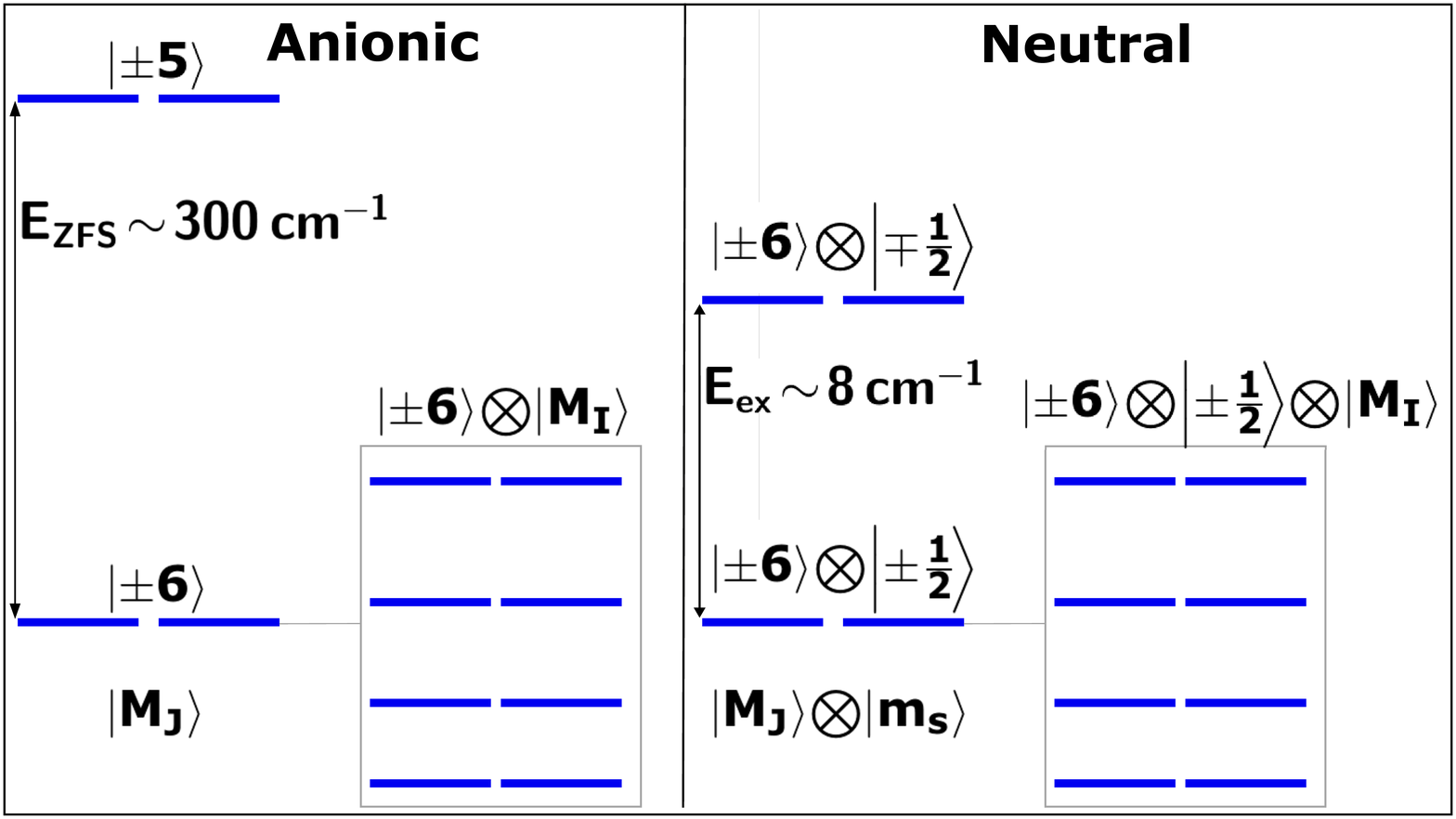}
\caption{Schematic illustration of the low-energy electronic-nuclear spectrum for the anionic and neutral TbPc$_2$ molecules. For the anionic molecule, the electronic ground quasi-doublet $\ket{M_J=\pm6}$ is well isolated from the first excited quasi-doublet ($E_{\text{ZFS}}\sim300$ cm$^{-1}$). For the neutral molecule, the electronic ground doublet $\ket{M_J\!=\!\pm6}\!\otimes\!\ket{m_s\!=\!\pm1/2}$ is separated by a gap of $\sim$8 cm$^{-1}$ from the first-excited doublet that differs by a relative orientation of the ligand spin and Tb angular momentum. For the anionic and neutral TbPc$_2$, the hyperfine interaction with the Tb nuclear spin $I=3/2$ splits each state of the electronic ground (quasi)doublet into four states. Here
$|M_I\rangle$ and $|m_s\rangle$ correspond to the nuclear spin and the ligand spin, respectively. Modified from Ref.\citenum{Pederson2019}.}
\label{Spectrum}
\end{figure}

The 4f$^8$ electronic configuration of the Tb$^{3+}$ ion results in electron spin and orbital angular momentum of $S=3$ and $L=3$, respectively. When the strong spin-orbit interaction (SOI) is combined with the ligand crystal field (CF), the Tb ion with the total angular momentum $J=6$ acquires large magnetic anisotropy and magnetic easy axis approximately perpendicular to the Pc planes. For cationic or anionic TbPc$_2$ SMMs, the ground $J=6$ multiplet is very well separated from excited multiplets. The lowest quasi-doublet ($M_J=\pm6$) in the ground $J$-multiplet is separated
from the first-excited quasi-doublet by $\sim$300~cm$^{-1}$ (Fig.~\ref{Spectrum}) \cite{Ungur2017,Pederson2019}, where $M_J$ is the total angular momentum projected onto the magnetic easy axis. For neutral TbPc$_2$ SMMs, one unpaired electron (with spin $s=1/2$) is delocalized within the two Pc ligands. This ligand spin interacts with the Tb multiplets and doubles the number of low-energy levels. In addition, this extra electron makes neutral TbPc$_2$ a Kramers system. The ground doublet can be represented by $M_J=\pm6$ coupled to the ligand spin parallel to the Tb angular momentum. The first-excited doublet lies $\sim8$~cm$^{-1}$ above and it can be represented by $M_J=\pm6$ coupled to the ligand spin anti-parallel to the Tb angular momentum (Fig.~\ref{Spectrum}).\cite{Pederson2019}

$^{159}$Tb nucleus has the natural abundance of 100\% and its nuclear spin is $I=3/2$. Therefore, the Tb nuclear spin couples to the electronic
spin and orbital angular momentum, which splits each electronic state into four states. Since $I > 1/2$, the nuclear quadrupole interaction must also be present in TbPc$_2$ SMMs. Taking the ground $J=6$ multiplet into account, observed magnetic hysteresis loops were fitted to a simplified
model with two parameters such as a magnetic hyperfine coupling constant and an axial nuclear quadrupole parameter \cite{Ishikawa2005}. However, there are no {\it ab-initio} studies on the nature and mechanism of the magnetic hyperfine and nuclear quadrupole
interactions in these SMMs. Density-functional theory cannot be used for such studies because the Tb$^{3+}$ ion has large orbital angular
momentum as well as almost degenerate $4f$ orbitals.

Here we investigate the magnetic hyperfine and nuclear quadrupole interactions for anionic and neutral TbPc$_2$ molecules using {\it ab-initio} multirefrence calculations including SOI. We evaluate the interaction parameters and analyze their physical and chemical origin. We then
construct the effective pseudo-spin Hamiltonian from which the electronic-nuclear low-energy spectrum is found. Our results unambiguously show
the necessity of using multireference methods including SOI for the proposed study. Next, we calculate the Zeeman diagram and discuss the role of the
magnetic hyperfine and nuclear quadrupole interactions in magnetization dynamics. Finally, we investigate the effect of molecular distortions
on these interactions.



\section{Methodology}

In the following we use SI units. Quantum-mechanical operators are denoted by hats. Vectors, tensors and matrices are written in a bold font.

\subsection{Magnetic hyperfine interactions}

Consider a non-relativistic electron with a mass $m$, a linear momentum $\hat{\mathbf{p}}$, and a spin ${\hat {\mathbf s}}$ in the presence of
a vector potential $\hat{\mathbf{A}}_{\text N}$ induced by a nuclear spin moment $\hat{\mathbf{I}}$ where the nucleus is located at the origin.
The microscopic Hamiltonian $\hat{H}_{\text{MHf}}$ for such an electron can be written as\cite{AbragamBook}
\begin{equation}
\hat{H}_\text{MHf}=\frac{1}{2m}\left(\hat{\mathbf{p}}+e\hat{\mathbf{A}}_\text{N}\right)^2 + g_\text{e}\mu_\text{B}(\nabla\times\hat{\mathbf{A}}_\text{N})\cdot\hat{\mathbf{s}},
\label{eq:HMHf-0}
\end{equation}
where $g_{\text e}$ is the electronic $g$-factor, $\mu_{\text B}$ is Bohr magneton, and $\hat{\mathbf{A}}_\text{N}=\hat{\mathbf{m}}_\text{N}\times\mathbf{r}/r^3$. Here $\hat{\mathbf{m}}_{\text N}=g_\text{N}\mu_\text{N}\hat{\mathbf{I}}$ denotes the nuclear magnetic moment, where $g_\text{N}$ is the nuclear $g$-factor (that is 1.34267 for Tb), $\mu_N$ is the nuclear magneton, and $\mathbf{r}$ is the position vector of the electron with respect to the nucleus. Keeping only the first-order terms in $\hat{\mathbf{A}}_{\text N}$, this Hamiltonian can be decomposed into three contributions such as
\begin{eqnarray}
\hat{H}_{\text{MHf}} &=& \hat{H}_{\text{PSO}}+\hat{H}_{\text{SD}}+\hat{H}_{\text{FC}}.
\label{eq:HMHf}
\end{eqnarray}
The first term in Eq.~(\ref{eq:HMHf}) is the paramagnetic spin-orbital (PSO) contribution that describes the interaction between the electronic
orbital angular momentum $\hat{\mathbf{L}}$ and the nuclear magnetic moment
\begin{equation}
\hat{H}_{\text{PSO}}=\frac{\mu_0\mu_B}{2\pi}\frac{1}{r^3}\hat{\mathbf{L}}\cdot\hat{\mathbf{m}}_N,
\label{eq:HPSO}
\end{equation}
where $\mu_0$ is the vacuum permeability. The second term in Eq.~(\ref{eq:HMHf}) describes the spin dipolar (SD) interaction between the nuclear magnetic moment and the electronic spin moment $\hat{\mathbf{m}}_s=g_{\text e}\mu_B\hat{\mathbf{s}}$
\begin{equation}
\hat{H}_{\text{SD}}=\frac{\mu_0}{4\pi}\frac{1}{r^3}\bigg[\hat{\mathbf{m}}_s\cdot\hat{\mathbf{m}}_N
-3\frac{(\hat{\mathbf{m}}_s\cdot{\mathbf{r}})(\hat{\mathbf{m}}_N\cdot{\mathbf{r}})}{r^2}\bigg].
\label{eq:HSD}
\end{equation}
Finally, $\hat{H}_{\text{FC}}$ represents the Fermi contact (FC) interaction between the spin density at the nucleus position and the nuclear
magnetic moment
\begin{equation}
\hat{H}_{\text{FC}}=-\frac{\mu_0}{4\pi}\frac{8\pi}{3}\hat{\mathbf{m}}_s\cdot\hat{\mathbf{m}}_N\delta({\mathbf{r}}),
\label{eq:HFC}
\end{equation}
where $\delta({\mathbf r})$ is a three-dimensional Dirac delta function. Note that for a Tb$^{3+}$ ion with large orbital angular momentum,
the non-relativistic form of magnetic hyperfine operator, Eqs.~(\ref{eq:HMHf}-\ref{eq:HFC}), is a good approximation.\cite{Bolvin2014}

For both the anionic and neutral TbPc$_2$ SMMs, the energy gap between the electronic ground and excited (quasi)doublets is much greater
than the scale of the magnetic hyperfine interaction ($\sim0.1$ cm$^{-1}$). Therefore, for the low-energy electronic-nuclear spectrum
only the ground (quasi)doublet (Fig.~\ref{Spectrum}) is relevant and it can be represented by a fictitious pseudo-spin $S=1/2$. 
The pseudo-spin then interacts with the magnetic field produced by the Tb nuclear spin $I=3/2$. The effective pseudo-spin Hamiltonian 
for this magnetic hyperfine interaction can be written as\cite{AbragamBook}
\begin{equation}
\hat{H}_{A}=\hat{\mathbf{I}}\cdot\mathbf{A}\cdot\hat{\mathbf{S}},
\label{eq:HA}
\end{equation}
where $\mathbf{A}$ is the magnetic hyperfine matrix. In order to understand degrees of mixing among $\ket{M_S,M_I}$ levels,
it is convenient to rewrite the above using the ladder operators
\begin{equation}
\hat{H}_{A}=\frac{A_{zz}}{2}\hat{I}_z\hat{S}_z + \frac{A_0}{2}\hat{I}_+\hat{S}_- +A_1\left(\hat{I}_-\hat{S}_z+\hat{I}_z\hat{S}_-\right) +\frac{A_2}{2}\hat{I}_-\hat{S}_- + \text{h.c.},
\label{eq:HA2}
\end{equation}
where h.c. denotes Hermitian conjugated terms. Here, we introduce complex parameters that are combinations of elements of the
$\mathbf{A}$ matrix
\begin{eqnarray}
A_0=\frac{1}{2}\left(A_{xx}+A_{yy}\right), \\
A_1=\frac{1}{2}\left(A_{xz}+iA_{yz}\right), \\
A_2=\frac{1}{2}\left(A_{xx}-A_{yy}\right)+iA_{xy}.
\end{eqnarray}

For pseudo-spin $S=1/2$, the eigenvalues of the effective pseudo-spin Hamiltonian, Eq.~(\ref{eq:HA}), are analytically known \cite{AbragamBook},
and so they can be mapped onto the eigenvalues of the microscopic Hamiltonian, Eq.~(\ref{eq:HMHf}), obtained from the {\it ab initio}
methods, in order to evaluate the $\mathbf{A}$ matrix \cite{Bolvin2014,Sharkas2015}. We can determine only the magnitude of the matrix elements
because the eigenvalues of Eq.~(\ref{eq:HA}) are expressed in terms of the elements of $(\mathbf{A}\mathbf{A}^T)$ tensor:
\begin{equation}
(\mathbf{A}\mathbf{A}^T)_{\alpha\beta}=2\sum_{ij}\langle i|\hat{h}_{\text{MHf}}^\alpha|j\rangle\langle j|\hat{h}_{\text{MHf}}^\beta|i\rangle,
\label{eq:Aformula}
\end{equation}
where $\alpha, \beta=x,y,z$ and $\hat{h}_{\text{MHf}}^\alpha\equiv\partial\hat{H}_{\text{MHf}}/\partial\hat{I}_\alpha$.
The summation runs over the {\it ab initio} electronic ground (quasi)doublet ($i=1,2$).

\subsection{Nuclear quadrupole interactions}

A nucleus with $I>1/2$ has a nonzero quadrupole moment which interacts with an electric-field gradient at the nucleus position. The electric-field gradient operator $\hat{V}_{\alpha\beta}$ is given by $\frac{\partial^2\Phi}{\partial x_\alpha \partial x_\beta}$~\cite{Autschbach2010},
where $\Phi$ is the electrostatic potential produced by the electronic charge density and other nuclei at the position of the nucleus of interest.
The quadrupole interaction can be described by the following effective Hamiltonian\cite{AbragamBook}:
\begin{equation}
\hat{H}_{\text Q}=\hat{\mathbf{I}}\cdot\mathbf{P}\cdot\hat{\mathbf{I}}
\label{eq:HQ}
\end{equation}
where $\mathbf{P}$ is the nuclear quadrupole tensor. The components of $\mathbf{P}$ are related to the expectation value of
$\hat{V}_{\alpha\beta}$ over the {\it ab-initio} electronic states by the following:
\begin{equation}
P_{\alpha\beta}=\frac{Q}{2I(2I-1)}\langle\hat{V}_{\alpha\beta}\rangle,
\label{eq:Pmatrix}
\end{equation}
where $Q$ is the quadrupole constant that for Tb nucleus is 1432.8~mbarn.\cite{Pyykko2008} In order to understand the amount of mixing among
different $|M_I\rangle$ levels, we rewrite Eq.~(\ref{eq:HQ}) using the ladder operators such as
\begin{equation}
\hat{H}_Q=\frac{3}{4}P_{zz}\left[\hat{I}_z^2-\frac{I(I+1)}{3}\right]+P_1\left(\hat{I}_z\hat{I}_-+\hat{I}_-\hat{I}_z\right)+\frac{P_2}{2}\hat{I}_-^2 +\text{h.c.},
\label{eq:HQ2}
\end{equation}
where complex quadrupole parameters are
\begin{eqnarray}
P_1 &=& P_{xz}+iP_{yz}, \\
P_2 &=& \frac{1}{2}\left(P_{xx}-P_{yy}\right)+iP_{xy}.
\end{eqnarray}

\section{Computational details}

The \emph{ab initio} calculations are performed using the MOLCAS quantum chemistry code (version 8.2).\cite{Molcas} Scalar relativistic effects are included based on the Douglas-Kroll-Hess Hamiltonian\cite{Douglass1974,Hess1986} using relativistically contracted atomic natural orbital (ANO-RCC) basis sets.\cite{Widmark1990,Roos2004} In particular, polarized valence triple-$\zeta$ quality (ANO-RCC-VTZP) is used for the Tb ion, and polarized valence double-$\zeta$ quality (ANO-RCC-VDZP) is used for the nitrogen and carbon atoms. Valence double-$\zeta$ quality (ANO-RCC-VDZ) is used for the hydrogen atoms. This choice of the basis set was shown to produce an accurate description of the low-energy electronic levels of TbPc$_2$-type molecules.\cite{Pederson2019}

Electronic structure is calculated in two steps. In the first step, in the absence of SOI, for a given spin multiplicity, the spin-free eigenstates are obtained using state-averaged complete active space self-consistent field (SA-CASSCF) method.\cite{Roos1980,Siegbahn1981} For the anionic TbPc$_2$, the active space consists of eight electrons and seven 4$f$-like orbitals. For the neutral TbPc$_2$, an additional electron and a ligand orbital are included in the active space. We check that inclusion of extra ligand orbitals in the active space does not significantly affect calculated magnetic hyperfine and nuclear quadrupole interaction parameters (see Tables S1 and S2 in the Supporting Information).
For the anionic TbPc$_2$, we consider only $S=3$ configuration that corresponds to the Tb$^{3+}$ spin. For the neutral TbPc$_2$, depending on whether ligand electron spin is parallel or antiparallel to the Tb$^{3+}$ spin, we have two possible values of the total spin of the molecule: $S=7/2$ or $S=5/2$. Both spin configurations are included in the CASSSCF calculations because they lie close in energy.
For a given spin configuration in both the anionic and neutral molecules, we evaluate seven lowest spin-free states (roots) that roughly correspond to seven configurations of eight electrons in Tb $4f$-type orbitals. These seven spin-free states are used in the CASSCF state-averaged procedure.
In the second step, SOI is included within the atomic mean-field approximation,\cite{Hess1996} in the aforementioned spin configurations and spin-free eigenstates, using the restricted active space state-interaction (RASSI) method.\cite{rassi} The resulting electronic structure agrees well with previous works.\cite{Pederson2019,Ungur2017}

Having obtained the electronic structure, we calculate the $\mathbf{A}$ matrix according to Eq.~(\ref{eq:Aformula}) by evaluating the matrix elements of PSO, SD, and FC contributions within the ground (quasi)doublet. In MOLCAS version 8.2, only FC and SD contributions are included.\cite{Sharkas2015} Therefore, we implement the PSO term in the MOLCAS code. The main ingredient of the implementation is the evaluation of the one-electron integrals
for the $\hat{\mathbf{L}}/{r}^3$ operator. This is done by modification of the {\tt AMFI} module. We test our implementation by calculating the $\mathbf{A}$ matrix for CN, NpF$_6$ and UF$_6^-$ molecules (Table S3 in Supporting Information). As shown, our calculations agree well with the literature~\cite{Lan2014,Sharkas2015}.
The nuclear quadrupole tensor elements are calculated using Eq.~(\ref{eq:Pmatrix}) by evaluating the matrix elements of $\hat{V}_{\alpha\beta}$
between the ground (quasi)doublet. Note that this capability already exists in MOLCAS version 8.2.


\section{Results and Discussion}


We consider two TbPc$_2$ molecules: (1) neutral TbPc$_2$ with experimental geometry from Ref.~\citenum{Komijani2018} (Fig.~\ref{Geometry}a)
and (2) anionic TbPc$_2$ with experimental geometry from Ref.~\citenum{Branzoli2009} (Fig.~\ref{Geometry}b). The structural deviations from
$D_{4h}$ symmetry are much stronger for the anionic molecule. The ground state of the neutral molecule is a Kramers doublet which can be
characterized by $M_J=\pm6$ and the ligand spin parallel to the Tb angular momentum. On the other hand, for the anionic molecule, the ground
state is $M_J=\pm6$ quasidoublet with tunnel splitting of $\sim$140 MHz. For both molecules, the electronic (quasi)doublet can be
represented by a fictitious pseudo-spin $S=1/2$.

It is convenient to present the calculated $\mathbf{A}$ matrix and $\mathbf{P}$ tensor in the magnetic coordinate system (Fig.~\ref{Geometry})
in which the $\mathbf{g}$ matrix for the ground (quasi)doublet is diagonal. The $\mathbf{g}$ matrix has only one nonzero eigenvalue for the
anionic TbPc$_2$ molecule. This is expected because the ground quasi-doublet is not a Kramers system~\cite{Griffith1963}.
For the neutral TbPc$_2$ molecule, the ${\mathbf{g}}$ matrix has only one large eigenvalue with very small two other eigenvalues ($\sim$10$^{-6}$),
which can be explained by the fact that the ground doublet is well separated from excited doublets.
We choose the $z$ axis to point along the eigenvector corresponding to the large eigenvalue. This direction points approximately perpendicular to
the ligand planes. For the anionic TbPc$_2$, we find $g_{zz}=17.993$. This value agrees well with $2g_LJ=18$ (where $g_L=3/2$ is the Lande $g$
factor for the $S=3$, $L=3$, $J=6$ multiplet) as expected for the $M_J=\pm J$ doublet. For the neutral TbPc$_2$, we obtain $g_z=20.003$
which is larger by about two units than the value for the anionic molecule. This difference can be explained by the electronic $g$ factor of the
ligand electron. Note that since $g_{xx}=g_{yy}=0$, the $x$ and $y$ directions are not well defined. Therefore, it is more convenient to use
the complex parameters introduced in Eqs.~(\ref{eq:HA2}) and (\ref{eq:HQ2}).

\subsection{Magnetic hyperfine interaction}

\begin{table}[t!]
\centering
\caption{Calculated Elements\textsuperscript{\emph{a}} of the Magnetic Hyperfine Matrix in Units of MHz}
\begin{tabular}{l|cccccc|ccc}
\hline
 Molecule & $A_{xx}$ & $A_{yy}$ & $A_{zz}$ & $A_{xy}$ & $A_{xz}$ & $A_{yz}$ & $|A_0|$ & $|A_1|$ & $|A_2|$ \\
\hline
 Neutral  & 0.0 & 0.0 & 6145.8 & 0.0 & 0.8 &  0.0 & 0.0 & 0.4 & 0.0 \\
 Anionic  & 0.0 & 0.0 & 5992.6 & 0.0 & 0.1 & -0.5 & 0.0 & 0.3 & 0.0 \\
 \hline
\end{tabular}
\\
\raggedright
\textsuperscript{\emph{a}} Used the magnetic coordinate system (Fig.~\ref{Geometry}) in which the ${\mathbf g}$ matrix for the electronic
ground doublet is diagonal.
\label{AMatrix}
\end{table}

Table~\ref{AMatrix} shows the calculated elements of the magnetic hyperfine matrix. For both the neutral and anionic molecules, the $A_{zz}$ element 
is dominant with the other elements being close to zero. This is expected for large uniaxial magnetic anisotropy. For the entire $J=6$ 
ground-multiplet the ${\mathbf A}$ matrix is isotropic as long as interactions with higher multiplets can be neglected.\cite{AbragamBook} Projection on the ground doublet (with a pseudo-spin $S=1/2$) would then result in the interaction of the form $A_{zz}I_zS_z$. This is, indeed, the case as long as we use coordinate system in which the ${\mathbf A}$ matrix is diagonal. The presence of nonzero $A_{xz}$ and/or $A_{yz}$ elements is due to a slight misalignment between the $z$ axes of the $\mathbf{g}$ matrix and $\mathbf{A}$ matrix coordinate systems. However, $A_{xx}$, $A_{xy}$, and $A_{yy}$ terms still remain zero because $g_{xx}=g_{yy}=0$. For the anionic molecule, this misalignment is caused
by the interaction of the $J=6$ ground-multiplet with higher multiplets.\cite{AbragamBook} In the case of the neutral molecule, the interaction between the unpaired ligand spin and the ground-multiplet also contributes to this misalignment.
Note that the sign of the $A_{zz}$ is undetermined in our calculations.

\begin{figure}[t!]
\centering
\includegraphics[width=1.0\linewidth]{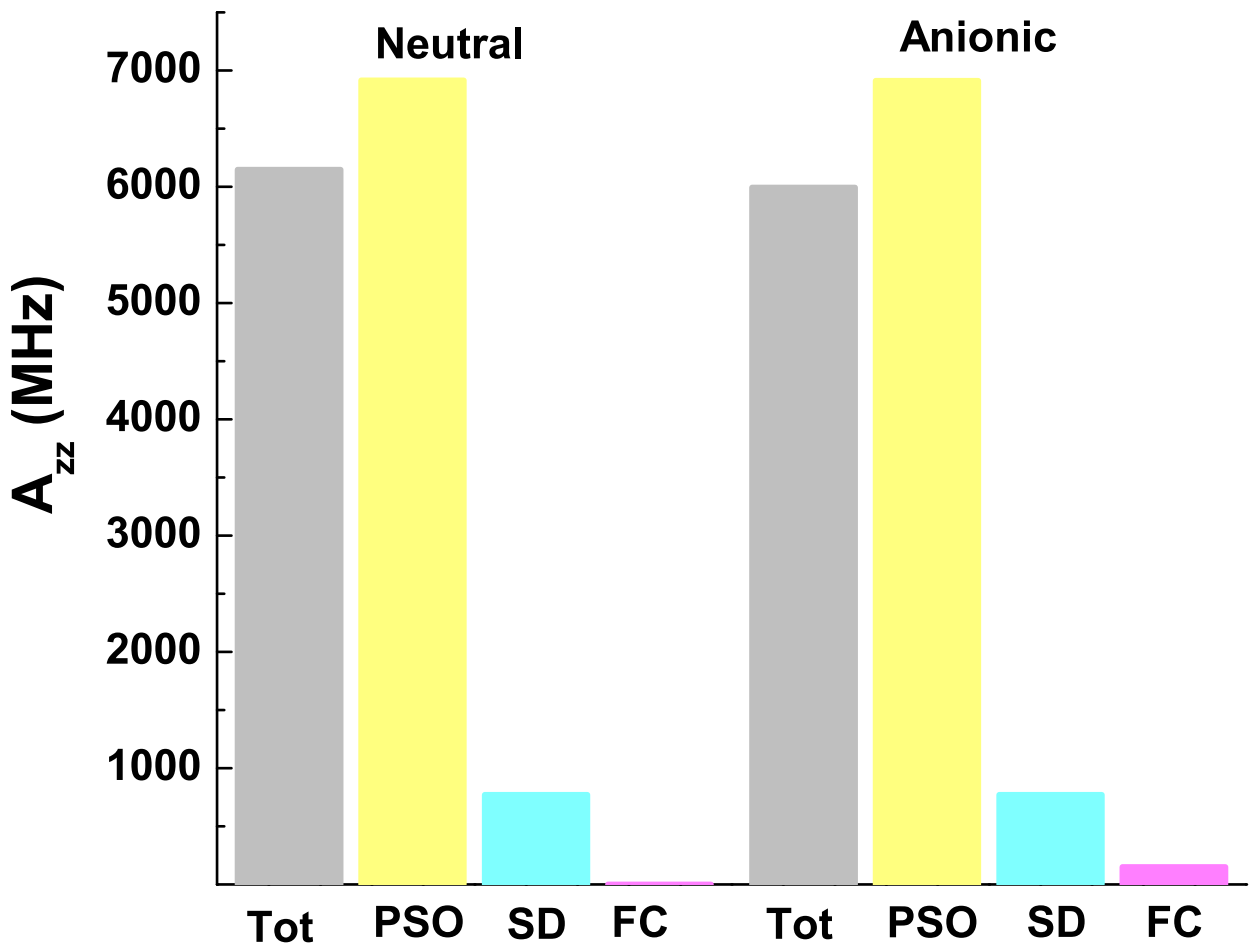}
\caption{Total (Tot), paramagnetic spin-orbital (PSO), spin-dipole (SD), and Fermi contact (FC) contributions to the calculated $A_{zz}$ element for the neutral and anionic TbPc$_2$ molecules. For the neutral molecule, the FC contribution is less than 0.1 MHz.}
\label{MagHyperfine}
\end{figure}

Overall, the calculated $A_{zz}$ value is about 6000 MHz. In order to understand the origin of such strong hyperfine coupling, we compare PSO, SD and FC contributions to $A_{zz}$ against its total value. See Figure~\ref{MagHyperfine}. For both molecules, the PSO contribution is dominant due to the presence of the large Tb orbital angular momentum. Both the SD and FC terms have opposite contributions to the PSO term. The SD contribution
is much smaller than the PSO part but still significant. The PSO and SD contributions are similar for the neutral and anionic molecules. However, the FC term differs strongly between different TbPc$_2$ molecules.
For the neutral molecule, the FC contribution is negligibly small ($<0.1$ MHz). On the other hand, for the anionic molecule, the FC contribution is $\approx$150 MHz. The fact that the FC is smaller than the PSO and SD contributions is not surprising, since the spin density is carried by $4f$-like orbitals which have zero spin density at the Tb nucleus position. The FC contribution requires hybridization between $4f$-like and $s$-like orbitals. Such hybridization would be stronger for the less symmetric anionic molecule, which likely explains a significantly larger FC term for this system. Note that the core-polarization effects are not included in our calculations, and therefore, the FC is expected to be somewhat underestimated. Nevertheless, for Tb$^{3+}$ ion, this effect is expected to be much smaller than the PSO and SD terms.\cite{AbragamBook}

\subsection{Nuclear quadrupole interaction}

\begin{table}[t!]
\centering
\caption{Calculated Elements\textsuperscript{\emph{a}} of the Nuclear Quadrupole Tensor in Units of MHz for Neutral and Anionic TbPc$_2$ Molecules}
\begin{tabular}{l|cccccc|cc}
\hline
 Molecule & $P_{xx}$ & $P_{yy}$ & $P_{zz}$ & $P_{xy}$ & $P_{xz}$ & $P_{yz}$ & $|P_1|$ & $|P_2|$ \\
\hline
 Neutral  & -139.1 & -147.3 & 286.4 & -0.1 & 4.7 & -0.1  & 4.7  & 4.1  \\
 Anionic  & -54.8 & -217.2 & 272.0 & -28.4 & 9.2 & -16.0 & 18.5 & 86.0 \\
\hline
\end{tabular}
\\
\raggedright
\textsuperscript{\emph{a}} Used the magnetic coordinate system (Fig.~\ref{Geometry}) in which the ${\mathbf g}$ matrix for the electronic
ground doublet is diagonal.
\label{PMatrix}
\end{table}

Table~\ref{PMatrix} shows the calculated elements of the traceless nuclear quadrupole tensor. We obtain a strong quadrupole interaction with $P_{zz}$
approaching 300 MHz. This is a result of orbital angular momentum carrying asymmetric $4f$ charge distribution which creates a large electric-field gradient at the Tb nucleus. Indeed, we check that the replacement of the Tb ion by a Gd$^{3+}$ ion ($L=0$) results in a much smaller 
electric-field gradient. The diagonal $P_{zz}$ parameter is similar for the neutral and anionic TbPc$_2$. However, the off-diagonal elements ($P_1$ and $P_2$) differ significantly between the two TbPc$_2$ molecules due to different degrees of the structural deviations from
the $D_{4d}$ symmetry. For the neutral TbPc$_2$, the magnitudes of $P_1$ and $P_2$ are less than 5 MHz. On the other hand, for the anionic
TbPc$_2$, these parameters are significantly larger with $|P_2|$ being almost third of the $P_{zz}$ value.

\begin{figure}[t!]
\centering
\includegraphics[width=1.0\linewidth]{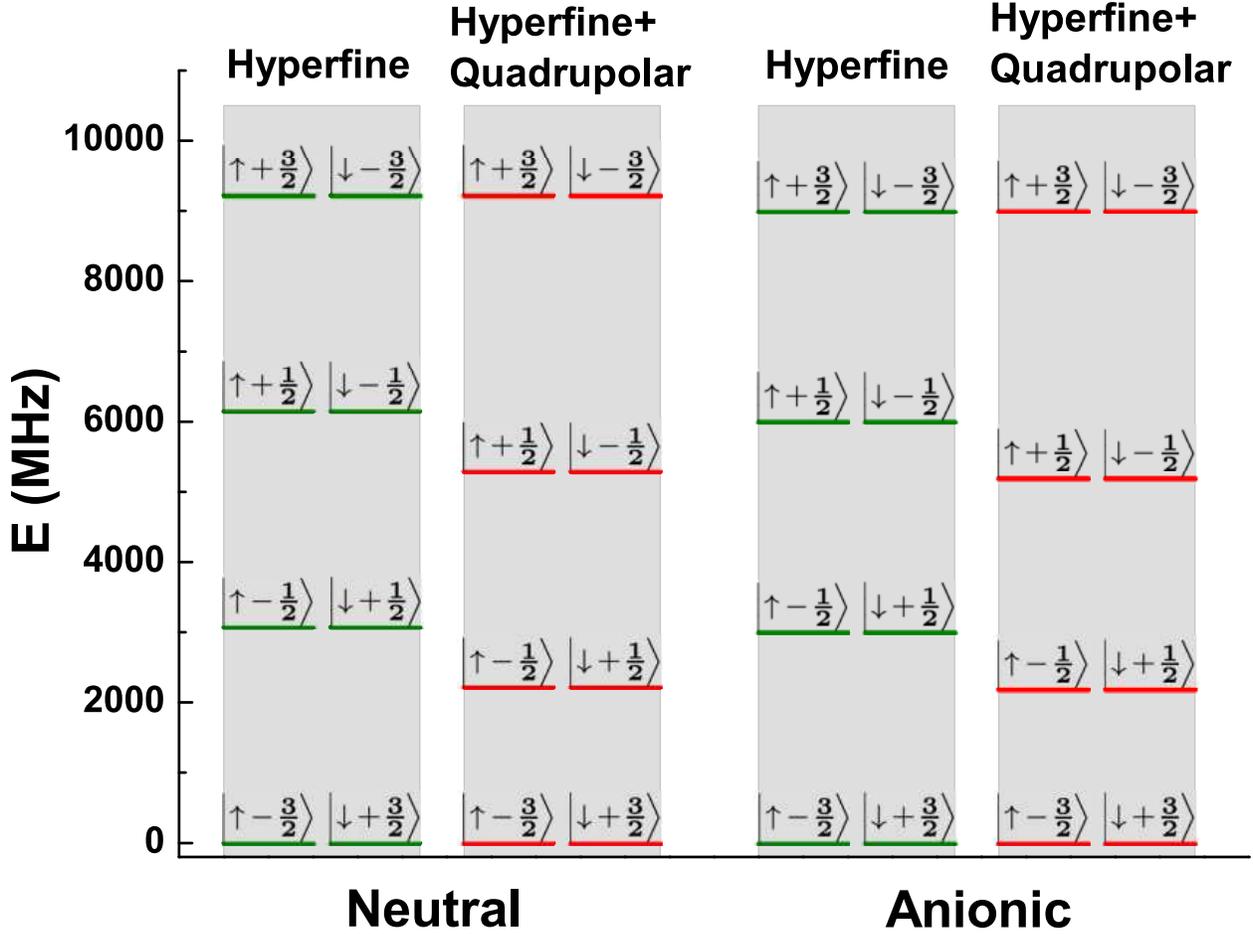}
\caption{The calculated low-energy electronic-nuclear spectra of the neutral and anionic TbPc$_2$ molecules. The equally spaced green lines correspond to energy levels found by diagonalization of the magnetic hyperfine Hamiltonian, Eq.~(\ref{eq:HA}). The red lines correspond to energy levels found by diagonalization of the sum of the magnetic hyperfine and nuclear quadrupole Hamiltonians [Eqs.~(\ref{eq:HA}) and (\ref{eq:HQ})]. The approximate quantum numbers $\ket{M_S,M_I}$ of the calculated energy levels are shown.}
\label{EnergyLevels}
\end{figure}

\subsection{Electronic-nuclear energy spectrum}

Having calculated the elements of the $\mathbf{A}$ matrix and $\mathbf{P}$ tensor for the electronic ground doublet, we now calculate the low-energy electronic-nuclear spectrum of TbPc$_2$. Figure~\ref{EnergyLevels} shows the energy levels calculated by diagonalizing the effective pseudo-spin Hamiltonian for the magnetic hyperfine interaction [Eq.~(\ref{eq:HA})] with and without the nuclear quadrupole interaction [Eq.~(\ref{eq:HQ})].
For both the neutral and anionic TbPc$_2$, the spectrum is composed of four doublets. This degeneracy is a consequence of the fact that the $\mathbf{A}$ matrix has only one nonzero eigenvalue. Therefore, at zero magnetic field, the Hamiltonian matrix is block-diagonal with two blocks,
one for each $S_z$ value. Within each block, the Hamiltonian matrix describes a $I=3/2$ system which must be at least doubly degenerate by virtue of the Kramers theorem.

Without the nuclear quadrupole term, the doublets are equidistant with level spacing $\sim$3000 MHz. In this case, the pseudo-spin ($M_S=\uparrow,\downarrow$) and nuclear ($M_I=\pm3/2,\pm1/2$) magnetic quantum numbers represent the electronic-nuclear states (Fig.~\ref{EnergyLevels}) since all elements of the $\mathbf{A}$ matrix except $A_{zz}$ are negligible. For the lowest electronic-nuclear doublet the electronic pseudo-spin is antiparallel to the nuclear spin due to our choice of a positive $A_{zz}$ element.
Addition of the quadrupole interaction makes the energy levels non-equidistant in agreement with experiment.\cite{Vincent2012,Thiele2014,Ishikawa2005} Table~\ref{Exp} compares the experimental relative energy levels~\cite{Thiele2014} with our calculated values for the neutral and anionic TbPc$_2$ molecules. Note that this approach is favorable over direct comparison between the theoretical and experimental interaction parameters. In the fitting to the experimental data, a less general Hamiltonian with only two parameters was used.\cite{Vincent2012,Thiele2014,Ishikawa2005} The calculated energy levels are similar for the neutral and anionic molecules. In both cases, the calculated values agree well with experiment. This indicates that our methodology is capable of an accurate quantitative description of the electronic-nuclear spectrum of Tb-based SMMs.

\begin{table}[t!]
\centering
\caption{Calculated and Experimental\textsuperscript{\emph{a}} Electronic-Nuclear Relative Energy Levels in GHz}
\begin{tabular}{l|cccccc|cc}
\hline
 Levels\textsuperscript{\emph{b}} & Neutral & Anionic & Experiment \\
\hline
 $E_2-E_1$  & 2.214 & 2.182 & 2.(5) \\
 $E_3-E_2$  & 3.073 & 3.003 & 3.(1) \\
 $E_4-E_3$  & 3.932 & 3.812 & 3.(7) \\
\hline
\end{tabular}
\\
\raggedright
\textsuperscript{\emph{a}} Ref. \citenum{Thiele2014}. \textsuperscript{\emph{b}} $E_i$ denotes $i^{\text{th}}$ lowest electronic-nuclear doublet.
\label{Exp}
\end{table}

The spectrum of the anionic and the neutral molecules are similar even though the two molecules have significantly different off-diagonal $\mathbf{P}$ tensor elements. This indicates that the transverse quadrupole parameters do not have a large effect on the energy levels. These parameters, however, mix states with different $M_I$. As a result, the electronic-nuclear states are not pure $\ket{M_S,M_I}$ states and the notation in Fig.~\ref{EnergyLevels} is only approximate. The mixing is illustrated in the Supporting Information (Tables~S4 and S5) where the eigenvectors of the effective pseudo-spin Hamiltonian are shown for the neutral and anionic TbPc$_2$, respectively. The mixing is significantly stronger for the asymmetric anionic molecule with larger off-diagonal $\mathbf{P}$-tensor elements. The mixing plays an important role in magnetization dynamics (see below).

\subsection{Zeeman diagram}

\begin{figure}[t!]
\centering
\includegraphics[width=1.0\linewidth]{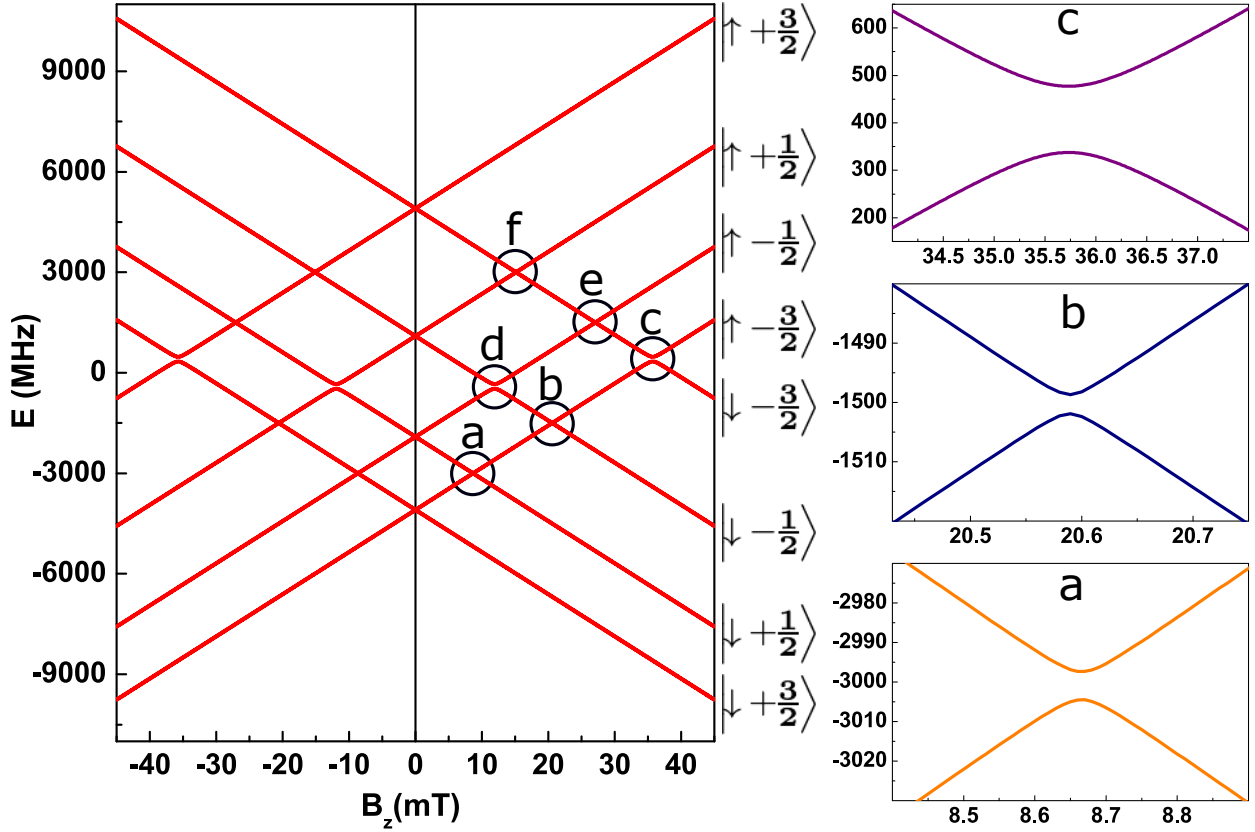}
\caption{(Left) Zeeman diagram showing the calculated electronic-nuclear energy levels as a function of magnetic field along the $z$ axis ($B_z$) 
for the anionic TbPc$_2$ molecule. Here $\ket{M_S,M_I}$ represents the approximate quantum numbers of the levels away from avoided level crossing
points. Black circles denote avoided level crossing points for the positive magnetic fields. (Right) Zoom-ins of three avoided level crossings 
({\tt a}, {\tt b} and {\tt c}).}
\label{ZeemanDiagram}
\end{figure}

In order to investigate the role of hyperfine and quadrupole interactions on magnetization dynamics we calculate Zeeman diagram that shows the evolution of electronic-nuclear energy levels as a function of external magnetic field along the $z$ axis ($B_z$) for the anionic TbPc$_2$ molecule. The magnetic field is included by considering the Zeeman pseudo-spin Hamiltonian $\hat{H}_{\text Z}=\mu_{\text B}B_{z}g_{zz}\hat{S}_z$. Since the anionic molecule is an even-electron system, the ground quasi-doublet has a small tunnel splitting attributed to the transverse CF. This tunnel splitting is crucial for a description of magnetization dynamics because it mixes states with $M_S=\uparrow,\downarrow$ levels, and allows tunneling between different electronic-nuclear levels. For the pseudo-spin $S=1/2$, the transverse CF can be described by $\hat{H}_{\text{TS}}=\Delta_{\text{TS}}\hat{S}_x$ \cite{AbragamBook,Griffith1963}, where $\Delta_{\text{TS}}$ is the tunnel splitting i.e., the
energy splitting between the lowest electronic quasi-doublet. The value of $\Delta_{\text{TS}}$ depends crucially on the details of TbPc$_2$ structure.\cite{Pederson2019} In particular, for the asymmetric anionic TbPc$_2$ of interest, $\Delta_{\text{TS}}$ is as large as $\sim$140 MHz.
The full effective pseudo-spin Hamiltonian is, thus,
\begin{equation}
\hat{H}_{\text{eff}}=\hat{H}_{\text A}+\hat{H}_{\text Q}+\hat{H}_{\text Z}+\hat{H}_{\text{TS}}.
\label{Heff}
\end{equation}

\begin{table}[t!]
\centering
\caption{Characteristics of Avoided Level Crossings for Positive Magnetic Fields for the Anionic TbPc$_2$ Molecule}
\begin{tabular}{l|ccc}
\hline
   & $B_{\text{ALC}}$ (mT) & $\Delta E_{\text{ALC}}$ (MHz) & $|\Delta M_{\text{ALC}}|$ \\
\hline
a  & 8.67            & 7        & 2          \\
b  & 20.59           & 3        & 1          \\
c  & 35.73           & 140      & 0          \\
d  & 11.93           & 140      & 0          \\
e  & 27.07           & 3        & 1          \\
f  & 15.14           & 7        & 2          \\
\hline
\end{tabular}
\\
\raggedright
\label{ALC}
\end{table}

Figure~\ref{ZeemanDiagram} shows the Zeeman diagram obtained using the parameters calculated for the anionic TbPc$_2$ molecule. The diagram is symmetric with respect to $B_z\rightarrow -B_z$. 
The external magnetic field breaks the zero-field two-fold degeneracy. At certain magnetic field values, the levels with opposite $M_S$ cross. For positive fields, these crossing points are denoted by black circles in Fig.~\ref{ZeemanDiagram}. Zooming the area around the crossing points reveals that the levels do not actually cross and there is a finite gap between them (see zoom-ins on the right of Fig.~\ref{ZeemanDiagram}). We have, thus, avoided level crossings (ALCs). At positive fields, there are six such ALCs. The magnetic field value $B_{\text{ALC}}$ and the width of the gap 
$\Delta E_{\text{ALC}}$ for each of these ALCs are shown in Table~\ref{ALC}. In addition, Table~\ref{ALC} provides the magnitude of the difference between $M_I$ of the two crossing levels $|\Delta M_{\text{AC}}|$. There are two ALCs with $|\Delta M_{\text{AC}}|=0$, two with $|\Delta M_{\text{ALC}}|=1$, and two with $|\Delta M_{\text{ALC}}|=2$. 
ALCs with the same $|\Delta M_{\text{ALC}}|$ have equal values of $\Delta E_{\text{ALC}}$. ALCs with $|\Delta M_{\text{ALC}}|=0$ have a significant gap that is roughly equal to $\Delta_{\text{TS}}$. Indeed, in this case the wave functions of the crossing levels can be approximated by pure $\ket{M_S,M_I}$ states and $\Delta E_{\text{ALC}}\approx2|\bra{\uparrow M_I}\hat{H}_{\text{eff}}\ket{\downarrow M_I}|=\Delta_{\text{TS}}$. However, for $|\Delta M_{\text{ALC}}|\neq0$, $\bra{\uparrow M_I}\hat{H}_{\text{eff}}\ket{\downarrow M_I'}=0$ and, therefore, we need to take into account the mixing between the states with different $M_I$. Using perturbation theory, it can be shown that for $|\Delta M_{\text{ALC}}|=1$, the ALC gap is controlled by terms like $\frac{\Delta_{\text{TS}}|P_1|}{E_2-E_1}$ and  $\frac{\Delta_{\text{TS}}|P_1|}{E_3-E_2}$. Similarly, for $|\Delta M_{\text{LAC}}|=2$, terms like $\frac{\Delta_{\text{TS}}|P_2|}{E_3-E_1}$ and $\frac{\Delta_{\text{TS}}|P_2|}{E_4-E_2}$ determine the ALC gap. Note that in this case the denominators are significantly larger but since, for the anionic TbPc$_2$, $|P_2|>|P_1|$, the corresponding $\Delta E_{\text{ALC}}$ is still larger than for $|\Delta M_{\text{ALC}}|=1$.


ALCs are essential for low-temperature magnetization dynamics of TbPc$_2$ SMMs. The ALC gap determines the tunneling probability between different electronic-nuclear levels as the magnetic field is swept with a fixed rate $|dB_z/dt|$. The probability is given by the Landau-Zener formula\cite{Wernsdorfer2000}
\begin{equation}
P_{\text{LZ}}=1-\text{exp}\left(-\frac{\pi\Delta E_{\text{ALC}}^2}{2\hbar\mu_B g_{zz}|dB_z/dt|} \right)
\end{equation}
Clearly, the tunneling probability is the largest for ALCs with the largest $\Delta E_{\text{ALC}}$ that is ALCs with $|\Delta M_{\text{ALC}}|=0$. 
The tunneling transitions with $|\Delta M_{\text{AC}}|\neq0$ can also occur as long as significant transverse quadrupole interaction parameters are present. Such transitions have recently been observed.\cite{Taran2019}


\subsection{Effect of geometry distortion}

\begin{figure}[t!]
\centering
\includegraphics[width=1.0\linewidth]{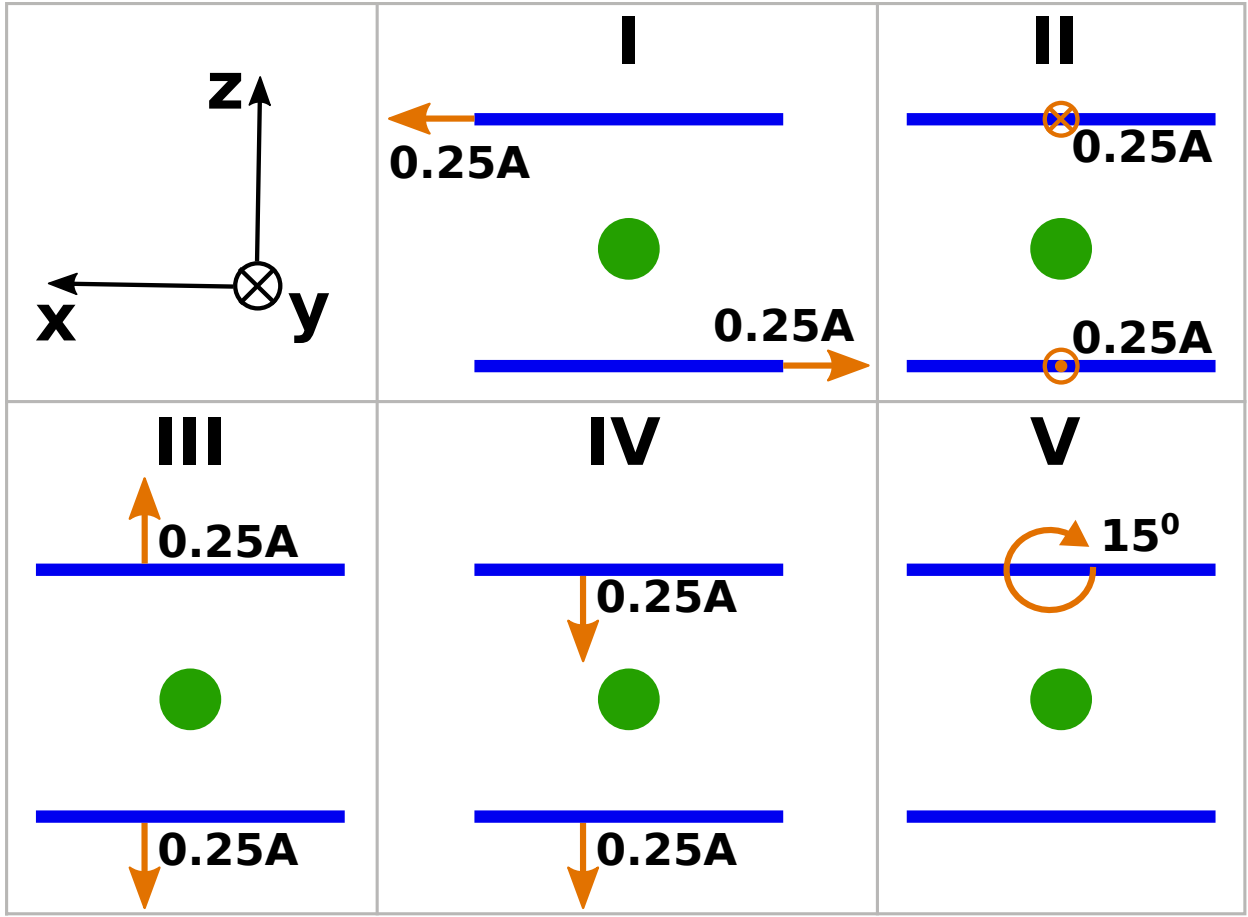}
\caption{Schematic illustration of five types of molecular distortions (denoted by roman letters) applied to the anionic TbPc$_2$. See the main text for the description of the distortions.}
\label{Distorted}
\end{figure}

In previous sections, we demonstrate that the transverse quadrupole parameters for the TbPc$_2$ molecule can vary significantly depending on differences in molecular structures. In order to investigate the effect of molecular distortion on hyperfine interactions in a more systematic manner we apply five types of ligand distortions to the anionic TbPc$_2$, as shown in Fig.~\ref{Distorted}: (I) Two Pc ligands displaced along the $x$ axis 
in the opposite direction to each other by 0.25~\AA, (II) Pc ligands displaced along the $y$ axis in the opposite direction to each other by 0.25~\AA, (III) Pc ligands displaced along the $z$ axis in the opposite direction to each other by 0.25~\AA, (IV) Pc ligands displaced along the $z$ axis in the same direction by 0.25~\AA, and (V) one of the Pc ligands rotated in the $xy$ plane by 15$^{\circ}$.

Table~\ref{Distortion} shows the hyperfine and quadrupole parameters, electronic-nuclear energy levels as well as tunnel splittings calculated for the five types of distortions of the anionic TbPc$_2$ molecule. We find that the PSO and SD contributions to $A_{zz}$ are insensitive to molecular distortions. On the other hand, the FC term shows significant variation under considered distortions which is primarily responsible for the changes of $A_{zz}$ in Table~\ref{Distortion}. In particular, the FC term increases for types I, II and III resulting in a decrease of the total $A_{zz}$ since FC term is opposite to the dominant PSO contribution. For types IV and V, the FC term is reduced and the total $A_{zz}$ value is enhanced. This result indicates that a potential route for an external control of the magnetic hyperfine interactions of the Tb nucleus in the Tb-based SMMs should involve modification of the FC contribution.

\begin{table}[t!]
\centering
\caption{Calculated Hyperfine and Quadrupole Parameters, Relative Electronic-Nuclear Energy Levels, and Tunnel Splitting for Different Distorted Geometries\textsuperscript{\emph{a}} of the Anionic TbPc$_2$ Molecule in Units of MHz}
\begin{tabular}{l|cccc|ccc|ccc|c}
\hline
                             & $A_{zz}$ & $|A_0|$ & $|A_1|$ & $|A_2|$ & $P_{zz}$ & $|P_1|$ & $|P_2|$ & $E_2\!-\!E_1$ & $E_3\!-\!E_2$ & $E_4\!-\!E_3$ & $\Delta_{\text{TS}}$\\
\hline
 I & 5951.0   & 0.0     & 0.6     & 0.0     & 240.3    & 127.8   & 82.5    & 2298    & 2948    & 3722     & 90 \\
 II & 5950.0   & 0.0     & 0.9     & 0.0     & 208.6    & 125.2   & 66.5    & 2389    & 2946    & 3626    & 90 \\
 III & 5873.1   & 0.0     & 0.3     & 0.0     & 311.0    & 21.8    & 104.7   & 2520    & 3460    & 4383    & 870 \\
 IV & 6082.7   & 0.0     & 0.1     & 0.0     & 203.7    & 8.4     & 60.0    & 2853    & 3467    & 4075    & 30  \\
 V & 6038.4   & 0.0     & 0.4     & 0.0     & 264.4    & 20.6    & 122.8   & 2229    & 3034    & 3811     & 180 \\
\hline
\end{tabular}
\\
\raggedright
\textsuperscript{\emph{a}} See Fig.~\ref{Distorted}.
\label{Distortion}
\end{table}

The diagonal quadrupole parameter $P_{zz}$ shows considerable variation under geometry distortions. This parameter is responsible for non-equidistant electronic nuclear spectrum which is crucial for an efficient readout of nuclear spin.\cite{Vincent2012,Thiele2014} For type III, the ligands are pulled away from the Tb ion reducing the hybridization between the 4$f$-like states and ligand orbitals. This increases the electric-field gradient at the Tb nucleus position and enhances $P_{zz}$. On the other hand, for types I, II and IV, the distortion increases hybridization of the Tb-ligand hybridization and reduces $P_{zz}$. Type V does not significantly affect the hybridization and, therefore, only a small change of $P_{zz}$ is found.

Transverse quadrupole parameters are strongly affected by geometry distortions. In particular, for types I and II, $|P_1|$ increases by an order of magnitude. As a result, the gap of $|\Delta M_{\text{ALC}}|=1$ ALCs is significantly enhanced (Table~S6 in the Supporting Information). This enhancement is, however, reduced by a decrease of $\Delta_{\text{TS}}$ under type I and II distortions (Table~\ref{Distortion}). In fact, using $\Delta_{\text{TS}}$ for the undistorted anionic TbPc$_2$ structure, we obtain even larger $|\Delta M_{\text{ALC}}|=1$ ALC gap for types I and II
distortions (Table~S7 in Supporting Information). Therefore, magnetization tunneling processes can significantly increase for certain structural distortions.

\section{Conclusions}

We investigate magnetic hyperfine and nuclear quadrupole interactions for $^{159}$Tb nucleus in TbPc$_2$ SMMs using multireference \emph{ab initio} calculations including SOI and the effective pseudo-spin Hamiltonian method. The key findings are as follows:

\begin{itemize}

\item The strong uniaxial magnetic hyperfine interaction ($A_{zz}\sim6000$ MHz) is dominated by the PSO mechanism.

\item Different experimental geometries have similar PSO and SD contributions to the magnetic hyperfine interaction, whereas the FC term shows a strong dependence on details of molecular geometry and becomes significant for less symmetric molecules.

\item Asymmetric 4$f$ charge distribution leads to a large diagonal quadrupole interaction ($P_{zz}\sim300$ MHz). For less symmetric molecular geometries, the transverse quadrupole parameters become significant.

\item The electronic-nuclear spectrum obtained by diagonalization of the effective pseudo-spin Hamiltonian is in excellent agreement with experiment.

\item The analysis of the calculated Zeeman diagram for the asymmetric anionic TbPc$_2$ molecule reveals that transverse quadrupole parameters allow for magnetization tunneling processes that do not conserve nuclear spin.

\item We demonstrate that small distortions of molecular geometry can affect hyperfine parameters with strong sensitivity of transverse quadrupole parameters to details of the molecular structure.

\end{itemize}

\begin{acknowledgement}
This work was funded by the Department of Energy (DOE) Basic Energy Sciences (BES) grant No DE-SC0018326. Computational support by Virginia Tech ARC and San Diego Supercomputer Center (SDSC) under DMR060009N. The authors are grateful to Dr. Kamal Sharkas for insightful discussion at the earlier
stage of this work.
\end{acknowledgement}

\begin{suppinfo}
The Supporting Information is available free of charge: Active space dependence, hyperfine calculations tests, pseudo-spin Hamiltonian eigenvectors, and avoided level crossing gaps for the distorted geometries.
\end{suppinfo}

\bibliography{refs}

\providecommand{\latin}[1]{#1}
\makeatletter
\providecommand{\doi}
  {\begingroup\let\do\@makeother\dospecials
  \catcode`\{=1 \catcode`\}=2 \doi@aux}
\providecommand{\doi@aux}[1]{\endgroup\texttt{#1}}
\makeatother
\providecommand*\mcitethebibliography{\thebibliography}
\csname @ifundefined\endcsname{endmcitethebibliography}
  {\let\endmcitethebibliography\endthebibliography}{}
\begin{mcitethebibliography}{47}
\providecommand*\natexlab[1]{#1}
\providecommand*\mciteSetBstSublistMode[1]{}
\providecommand*\mciteSetBstMaxWidthForm[2]{}
\providecommand*\mciteBstWouldAddEndPuncttrue
  {\def\EndOfBibitem{\unskip.}}
\providecommand*\mciteBstWouldAddEndPunctfalse
  {\let\EndOfBibitem\relax}
\providecommand*\mciteSetBstMidEndSepPunct[3]{}
\providecommand*\mciteSetBstSublistLabelBeginEnd[3]{}
\providecommand*\EndOfBibitem{}
\mciteSetBstSublistMode{f}
\mciteSetBstMaxWidthForm{subitem}{(\alph{mcitesubitemcount})}
\mciteSetBstSublistLabelBeginEnd
  {\mcitemaxwidthsubitemform\space}
  {\relax}
  {\relax}

\bibitem[Ses(2009)]{Sessoli2009}
Strategies Towards Single Molecule Magnets Based on Lanthanide Ions.
  \emph{Coord. Chem. Rev.} \textbf{2009}, \emph{253}, 2328 -- 2341\relax
\mciteBstWouldAddEndPuncttrue
\mciteSetBstMidEndSepPunct{\mcitedefaultmidpunct}
{\mcitedefaultendpunct}{\mcitedefaultseppunct}\relax
\EndOfBibitem
\bibitem[Baldov{\'{i}} \latin{et~al.}(2012)Baldov{\'{i}}, Cardona-Serra,
  Clemente-Juan, Coronado, Gaita-Ari{\~{e}}o, and Palii]{Baldovi2012}
Baldov{\'{i}},~J.~J.; Cardona-Serra,~S.; Clemente-Juan,~J.~M.; Coronado,~E.;
  Gaita-Ari{\~{e}}o,~A.; Palii,~A. Rational Design of Single-Ion Magnets and
  Spin Qubits Based on Mononuclear Lanthanoid Complexes. \emph{Inorg. Chem.}
  \textbf{2012}, \emph{51}, 12565--12574\relax
\mciteBstWouldAddEndPuncttrue
\mciteSetBstMidEndSepPunct{\mcitedefaultmidpunct}
{\mcitedefaultendpunct}{\mcitedefaultseppunct}\relax
\EndOfBibitem
\bibitem[Woodruff \latin{et~al.}(2013)Woodruff, Winpenny, and
  Layfield]{Woodruff2013}
Woodruff,~D.~N.; Winpenny,~R. E.~P.; Layfield,~R.~A. Lanthanide Single-Molecule
  Magnets. \emph{Chem. Rev.} \textbf{2013}, \emph{113}, 5110--5148\relax
\mciteBstWouldAddEndPuncttrue
\mciteSetBstMidEndSepPunct{\mcitedefaultmidpunct}
{\mcitedefaultendpunct}{\mcitedefaultseppunct}\relax
\EndOfBibitem
\bibitem[Liddle and van Slageren(2015)Liddle, and van Slageren]{Liddle2015}
Liddle,~S.~T.; van Slageren,~J. Improving f-Element Single Molecule Magnets.
  \emph{Chem. Soc. Rev.} \textbf{2015}, \emph{44}, 6655--6669\relax
\mciteBstWouldAddEndPuncttrue
\mciteSetBstMidEndSepPunct{\mcitedefaultmidpunct}
{\mcitedefaultendpunct}{\mcitedefaultseppunct}\relax
\EndOfBibitem
\bibitem[Wan(2016)]{Wang2016}
Single-Molecule Magnetism of Tetrapyrrole Lanthanide Compounds with Sandwich
  Multiple-Decker Structures. \emph{Coord. Chem. Rev.} \textbf{2016},
  \emph{306}, 195 -- 216\relax
\mciteBstWouldAddEndPuncttrue
\mciteSetBstMidEndSepPunct{\mcitedefaultmidpunct}
{\mcitedefaultendpunct}{\mcitedefaultseppunct}\relax
\EndOfBibitem
\bibitem[Guo \latin{et~al.}(2018)Guo, Day, Chen, Tong, Mansikkam{\"a}ki, and
  Layfield]{Guo2018}
Guo,~F.-S.; Day,~B.~M.; Chen,~Y.-C.; Tong,~M.-L.; Mansikkam{\"a}ki,~A.;
  Layfield,~R.~A. Magnetic Hysteresis up to 80 Kelvin in a Dysprosium
  Metallocene Single-Molecule Magnet. \emph{Science} \textbf{2018}, \emph{362},
  1400--1403\relax
\mciteBstWouldAddEndPuncttrue
\mciteSetBstMidEndSepPunct{\mcitedefaultmidpunct}
{\mcitedefaultendpunct}{\mcitedefaultseppunct}\relax
\EndOfBibitem
\bibitem[Goodwin \latin{et~al.}(2017)Goodwin, Ortu, Reta, Chilton, and
  Mills]{Goodwin2017}
Goodwin,~C.; Ortu,~F.; Reta,~D.; Chilton,~N.; Mills,~D. Molecular Magnetic
  Hysteresis at 60 Kelvin in Dysprosocenium. \emph{Nature} \textbf{2017},
  \emph{548}, 439--442\relax
\mciteBstWouldAddEndPuncttrue
\mciteSetBstMidEndSepPunct{\mcitedefaultmidpunct}
{\mcitedefaultendpunct}{\mcitedefaultseppunct}\relax
\EndOfBibitem
\bibitem[Thiele \latin{et~al.}(2014)Thiele, Balestro, Ballou, Klyatskaya,
  Ruben, and Wernsdorfer]{Thiele2014}
Thiele,~S.; Balestro,~F.; Ballou,~R.; Klyatskaya,~S.; Ruben,~M.;
  Wernsdorfer,~W. Electrically Driven Nuclear Spin Resonance in Single-Molecule
  Magnets. \emph{Science} \textbf{2014}, \emph{344}, 1135--1138\relax
\mciteBstWouldAddEndPuncttrue
\mciteSetBstMidEndSepPunct{\mcitedefaultmidpunct}
{\mcitedefaultendpunct}{\mcitedefaultseppunct}\relax
\EndOfBibitem
\bibitem[Godfrin \latin{et~al.}(2017)Godfrin, Ferhat, Ballou, Klyatskaya,
  Ruben, Wernsdorfer, and Balestro]{Godfrin2017}
Godfrin,~C.; Ferhat,~A.; Ballou,~R.; Klyatskaya,~S.; Ruben,~M.;
  Wernsdorfer,~W.; Balestro,~F. Operating Quantum States in Single Magnetic
  Molecules: Implementation of Grover's Quantum Algorithm. \emph{Phys. Rev.
  Lett.} \textbf{2017}, \emph{119}, 187702\relax
\mciteBstWouldAddEndPuncttrue
\mciteSetBstMidEndSepPunct{\mcitedefaultmidpunct}
{\mcitedefaultendpunct}{\mcitedefaultseppunct}\relax
\EndOfBibitem
\bibitem[Ishikawa \latin{et~al.}(2003)Ishikawa, Sugita, Ishikawa, Koshihara,
  and Kaizu]{Ishikawa2003}
Ishikawa,~N.; Sugita,~M.; Ishikawa,~T.; Koshihara,~S.-y.; Kaizu,~Y. Lanthanide
  Double-Decker Complexes Functioning as Magnets at the Single-Molecular Level.
  \emph{J. Am. Chem. Soc.} \textbf{2003}, \emph{125}, 8694--8695\relax
\mciteBstWouldAddEndPuncttrue
\mciteSetBstMidEndSepPunct{\mcitedefaultmidpunct}
{\mcitedefaultendpunct}{\mcitedefaultseppunct}\relax
\EndOfBibitem
\bibitem[He \latin{et~al.}(2014)He, Zhang, Hong, Cheng, Zhou, Shen, Li, Wang,
  Jiang, and Wu]{YangHe2014}
He,~Y.; Zhang,~Y.; Hong,~I.-P.; Cheng,~F.; Zhou,~X.; Shen,~Q.; Li,~J.;
  Wang,~Y.; Jiang,~J.; Wu,~K. Low-Temperature Scanning Tunneling Microscopy
  Study of Double-Decker DyPc$_2$ on Pb Surface. \emph{Nanoscale}
  \textbf{2014}, \emph{6}, 10779--10783\relax
\mciteBstWouldAddEndPuncttrue
\mciteSetBstMidEndSepPunct{\mcitedefaultmidpunct}
{\mcitedefaultendpunct}{\mcitedefaultseppunct}\relax
\EndOfBibitem
\bibitem[Wäckerlin \latin{et~al.}()Wäckerlin, Donati, Singha, Baltic,
  Rusponi, Diller, Patthey, Pivetta, Lan, Klyatskaya, Ruben, Brune, and
  Dreiser]{Wackerlin2016}
Wäckerlin,~C.; Donati,~F.; Singha,~A.; Baltic,~R.; Rusponi,~S.; Diller,~K.;
  Patthey,~F.; Pivetta,~M.; Lan,~Y.; Klyatskaya,~S.; Ruben,~M.; Brune,~H.;
  Dreiser,~J. Giant Hysteresis of Single-Molecule Magnets Adsorbed on a
  Nonmagnetic Insulator. \emph{Adv. Mater.} \emph{28}, 5195--5199\relax
\mciteBstWouldAddEndPuncttrue
\mciteSetBstMidEndSepPunct{\mcitedefaultmidpunct}
{\mcitedefaultendpunct}{\mcitedefaultseppunct}\relax
\EndOfBibitem
\bibitem[Studniarek \latin{et~al.}()Studniarek, Wäckerlin, Singha, Baltic,
  Diller, Donati, Rusponi, Brune, Lan, Klyatskaya, Ruben, Seitsonen, and
  Dreiser]{Studniarek2019}
Studniarek,~M.; Wäckerlin,~C.; Singha,~A.; Baltic,~R.; Diller,~K.; Donati,~F.;
  Rusponi,~S.; Brune,~H.; Lan,~Y.; Klyatskaya,~S.; Ruben,~M.; Seitsonen,~A.~P.;
  Dreiser,~J. Understanding the Superior Stability of Single-Molecule Magnets
  on an Oxide Film. \emph{Adv. Sci.} \emph{0}, 1901736\relax
\mciteBstWouldAddEndPuncttrue
\mciteSetBstMidEndSepPunct{\mcitedefaultmidpunct}
{\mcitedefaultendpunct}{\mcitedefaultseppunct}\relax
\EndOfBibitem
\bibitem[Ishikawa \latin{et~al.}(2004)Ishikawa, Sugita, Ishikawa, Koshihara,
  and Kaizu]{Ishikawa2004}
Ishikawa,~N.; Sugita,~M.; Ishikawa,~T.; Koshihara,~S.-y.; Kaizu,~Y. Mononuclear
  Lanthanide Complexes with a Long Magnetization Relaxation Time at High
  Temperatures: A New Category of Magnets at the Single-Molecular Level.
  \emph{J. Phys. Chem. B} \textbf{2004}, \emph{108}, 11265--11271\relax
\mciteBstWouldAddEndPuncttrue
\mciteSetBstMidEndSepPunct{\mcitedefaultmidpunct}
{\mcitedefaultendpunct}{\mcitedefaultseppunct}\relax
\EndOfBibitem
\bibitem[Vincent \latin{et~al.}(2012)Vincent, Klyatskaya, Ruben, Wernsdorfer,
  and Balestro]{Vincent2012}
Vincent,~R.; Klyatskaya,~S.~V.; Ruben,~M.; Wernsdorfer,~W.; Balestro,~F.
  Electronic Read-out of a Single Nuclear Spin Using a Molecular Spin
  Transistor. \emph{Nature} \textbf{2012}, \emph{488}, 357--360\relax
\mciteBstWouldAddEndPuncttrue
\mciteSetBstMidEndSepPunct{\mcitedefaultmidpunct}
{\mcitedefaultendpunct}{\mcitedefaultseppunct}\relax
\EndOfBibitem
\bibitem[Urdampilleta \latin{et~al.}(2013)Urdampilleta, Klyatskaya, Ruben, and
  Wernsdorfer]{Urdampilleta2013}
Urdampilleta,~M.; Klyatskaya,~S.; Ruben,~M.; Wernsdorfer,~W. Landau-Zener
  Tunneling of a Single Tb$^{3+}$ Magnetic Moment Allowing the Electronic
  Read-out of a Nuclear Spin. \emph{Phys. Rev. B} \textbf{2013}, \emph{87},
  195412\relax
\mciteBstWouldAddEndPuncttrue
\mciteSetBstMidEndSepPunct{\mcitedefaultmidpunct}
{\mcitedefaultendpunct}{\mcitedefaultseppunct}\relax
\EndOfBibitem
\bibitem[Thiele \latin{et~al.}(2013)Thiele, Vincent, Holzmann, Klyatskaya,
  Ruben, Balestro, and Wernsdorfer]{Thiele2013}
Thiele,~S.; Vincent,~R.; Holzmann,~M.; Klyatskaya,~S.; Ruben,~M.; Balestro,~F.;
  Wernsdorfer,~W. Electrical Readout of Individual Nuclear Spin Trajectories in
  a Single-Molecule Magnet Spin Transistor. \emph{Phys. Rev. Lett.}
  \textbf{2013}, \emph{111}, 037203\relax
\mciteBstWouldAddEndPuncttrue
\mciteSetBstMidEndSepPunct{\mcitedefaultmidpunct}
{\mcitedefaultendpunct}{\mcitedefaultseppunct}\relax
\EndOfBibitem
\bibitem[Godfrin \latin{et~al.}(2018)Godfrin, Ballou, Bonet, Ruben, Klyatskaya,
  Wernsdorfer, and Balestro]{Godfrin2018}
Godfrin,~C.; Ballou,~R.; Bonet,~E.; Ruben,~M.; Klyatskaya,~S.; Wernsdorfer,~W.;
  Balestro,~F. Generalized Ramsey Interferometry Explored with a Single Nuclear
  Spin Qudit. \emph{npj Quantum Inf.} \textbf{2018}, \emph{4}\relax
\mciteBstWouldAddEndPuncttrue
\mciteSetBstMidEndSepPunct{\mcitedefaultmidpunct}
{\mcitedefaultendpunct}{\mcitedefaultseppunct}\relax
\EndOfBibitem
\bibitem[Komijani \latin{et~al.}(2018)Komijani, Ghirri, Bonizzoni, Klyatskaya,
  Moreno-Pineda, Ruben, Soncini, Affronte, and Hill]{Komijani2018}
Komijani,~D.; Ghirri,~A.; Bonizzoni,~C.; Klyatskaya,~S.; Moreno-Pineda,~E.;
  Ruben,~M.; Soncini,~A.; Affronte,~M.; Hill,~S. Radical-Lanthanide
  Ferromagnetic Interaction in a $\mathrm{T}{\mathrm{b}}^{\mathrm{III}}$
  Bis-phthalocyaninato Complex. \emph{Phys. Rev. Materials} \textbf{2018},
  \emph{2}, 024405\relax
\mciteBstWouldAddEndPuncttrue
\mciteSetBstMidEndSepPunct{\mcitedefaultmidpunct}
{\mcitedefaultendpunct}{\mcitedefaultseppunct}\relax
\EndOfBibitem
\bibitem[Branzoli \latin{et~al.}(2009)Branzoli, Carretta, Filibian, Zoppellaro,
  Graf, Galan-Mascaros, Fuhr, Brink, and Ruben]{Branzoli2009}
Branzoli,~F.; Carretta,~P.; Filibian,~M.; Zoppellaro,~G.; Graf,~M.~J.;
  Galan-Mascaros,~J.~R.; Fuhr,~O.; Brink,~S.; Ruben,~M. Spin Dynamics in the
  Negatively Charged Terbium (III) Bis-phthalocyaninato Complex. \emph{J. Am.
  Chem. Soc.} \textbf{2009}, \emph{131}, 4387--4396\relax
\mciteBstWouldAddEndPuncttrue
\mciteSetBstMidEndSepPunct{\mcitedefaultmidpunct}
{\mcitedefaultendpunct}{\mcitedefaultseppunct}\relax
\EndOfBibitem
\bibitem[Ishikawa \latin{et~al.}(2004)Ishikawa, Sugita, Tanaka, Ishikawa,
  Koshihara, and Kaizu]{Ishikawa2004b}
Ishikawa,~N.; Sugita,~M.; Tanaka,~N.; Ishikawa,~T.; Koshihara,~S.; Kaizu,~Y.
  Upward Temperature Shift of the Intrinsic Phase Lag of the Magnetization of
  Bis(phthalocyaninato)terbium by Ligand Oxidation Creating an S = 1/2 Spin.
  \emph{Inorg. Chem.} \textbf{2004}, \emph{43}, 5498--5500\relax
\mciteBstWouldAddEndPuncttrue
\mciteSetBstMidEndSepPunct{\mcitedefaultmidpunct}
{\mcitedefaultendpunct}{\mcitedefaultseppunct}\relax
\EndOfBibitem
\bibitem[Loosli \latin{et~al.}(2006)Loosli, Liu, Neels, Labat, and
  Decurtins]{Loosli2006}
Loosli,~C.; Liu,~S.-X.; Neels,~A.; Labat,~G.; Decurtins,~S. Crystal Structures
  of Tetrabutylammonium Bis(phthalocyaninato)terbium(III) Methanol Solvate
  Hydrate [N(C$_4$H$_9$)$_4$][Tb(C$_8$H$_4$N$_2$)$_2$]·CH$_3$OH·3/2H$_2$O,
  and Tetrabutylammonium Bis(phthalocyaninato)dysprosium(III) Methanol Solvate
  Hydrate [N(C$_4$H$_9$)$_4$][Dy(C$_8$H$_4$N$_2$)$_2$]·CH$_3$OH·H$_2$O.
  \emph{Z. Kristallogr. Cryst. Mater} \textbf{2006}, \emph{221}, 135--141\relax
\mciteBstWouldAddEndPuncttrue
\mciteSetBstMidEndSepPunct{\mcitedefaultmidpunct}
{\mcitedefaultendpunct}{\mcitedefaultseppunct}\relax
\EndOfBibitem
\bibitem[Takamatsu \latin{et~al.}(2007)Takamatsu, Ishikawa, Koshihara, and
  Ishikawa]{Takamatsu2007}
Takamatsu,~S.; Ishikawa,~T.; Koshihara,~S.-y.; Ishikawa,~N. Significant
  Increase of the Barrier Energy for Magnetization Reversal of a
  Single-4$f$-Ionic Single-Molecule Magnet by a Longitudinal Contraction of the
  Coordination Space. \emph{Inorg. Chem.} \textbf{2007}, \emph{46},
  7250--7252\relax
\mciteBstWouldAddEndPuncttrue
\mciteSetBstMidEndSepPunct{\mcitedefaultmidpunct}
{\mcitedefaultendpunct}{\mcitedefaultseppunct}\relax
\EndOfBibitem
\bibitem[Katoh \latin{et~al.}(2009)Katoh, Yoshida, Yamashita, Miyasaka,
  Breedlove, Kajiwara, Takaishi, Ishikawa, Isshiki, Zhang, Komeda, Yamagishi,
  and Takeya]{Katoh2009}
Katoh,~K.; Yoshida,~Y.; Yamashita,~M.; Miyasaka,~H.; Breedlove,~B.~K.;
  Kajiwara,~T.; Takaishi,~S.; Ishikawa,~N.; Isshiki,~H.; Zhang,~Y.~F.;
  Komeda,~T.; Yamagishi,~M.; Takeya,~J. Direct Observation of
  Lanthanide(III)-Phthalocyanine Molecules on Au(111) by Using Scanning
  Tunneling Microscopy and Scanning Tunneling Spectroscopy and Thin-Film
  Field-Effect Transistor Properties of Tb(III)- and Dy(III)-Phthalocyanine
  Molecules. \emph{J. Am. Chem. Soc.} \textbf{2009}, \emph{131},
  9967--9976\relax
\mciteBstWouldAddEndPuncttrue
\mciteSetBstMidEndSepPunct{\mcitedefaultmidpunct}
{\mcitedefaultendpunct}{\mcitedefaultseppunct}\relax
\EndOfBibitem
\bibitem[Ganivet \latin{et~al.}(2013)Ganivet, Ballesteros, de~la Torre,
  Clemente-Juan, Coronado, and Torres]{Ganivet2013}
Ganivet,~C.~R.; Ballesteros,~B.; de~la Torre,~G.; Clemente-Juan,~J.~M.;
  Coronado,~E.; Torres,~T. Influence of Peripheral Substitution on the Magnetic
  Behavior of Single-Ion Magnets Based on Homo- and Heteroleptic TbIII
  Bis(phthalocyaninate). \emph{Chem. Eur. J} \textbf{2013}, \emph{19},
  1457--1465\relax
\mciteBstWouldAddEndPuncttrue
\mciteSetBstMidEndSepPunct{\mcitedefaultmidpunct}
{\mcitedefaultendpunct}{\mcitedefaultseppunct}\relax
\EndOfBibitem
\bibitem[Pederson \latin{et~al.}(2019)Pederson, Wysocki, Mayhall, and
  Park]{Pederson2019}
Pederson,~R.; Wysocki,~A.~L.; Mayhall,~N.; Park,~K. Multireference Ab Initio
  Studies of Magnetic Properties of Terbium-Based Single-Molecule Magnets.
  \emph{J. Phys. Chem. A} \textbf{2019}, \emph{123}, 6996--7006\relax
\mciteBstWouldAddEndPuncttrue
\mciteSetBstMidEndSepPunct{\mcitedefaultmidpunct}
{\mcitedefaultendpunct}{\mcitedefaultseppunct}\relax
\EndOfBibitem
\bibitem[Ungur and Chibotaru(2017)Ungur, and Chibotaru]{Ungur2017}
Ungur,~L.; Chibotaru,~L.~F. Ab Initio Crystal Field for Lanthanides.
  \emph{Chem. Eur. J} \textbf{2017}, \emph{23}, 3708--3718\relax
\mciteBstWouldAddEndPuncttrue
\mciteSetBstMidEndSepPunct{\mcitedefaultmidpunct}
{\mcitedefaultendpunct}{\mcitedefaultseppunct}\relax
\EndOfBibitem
\bibitem[Ishikawa \latin{et~al.}()Ishikawa, Sugita, and
  Wernsdorfer]{Ishikawa2005}
Ishikawa,~N.; Sugita,~M.; Wernsdorfer,~W. Quantum Tunneling of Magnetization in
  Lanthanide Single-Molecule Magnets: Bis(phthalocyaninato)terbium and
  Bis(phthalocyaninato)dysprosium Anions. \emph{Angew. Chem.} \emph{44},
  2931--2935\relax
\mciteBstWouldAddEndPuncttrue
\mciteSetBstMidEndSepPunct{\mcitedefaultmidpunct}
{\mcitedefaultendpunct}{\mcitedefaultseppunct}\relax
\EndOfBibitem
\bibitem[Abragam and Bleaney(1970)Abragam, and Bleaney]{AbragamBook}
Abragam,~A.; Bleaney,~B. \emph{Electron Paramagnetic Resonance of Transition
  Ions}; Clarendon Press: Oxford, 1970\relax
\mciteBstWouldAddEndPuncttrue
\mciteSetBstMidEndSepPunct{\mcitedefaultmidpunct}
{\mcitedefaultendpunct}{\mcitedefaultseppunct}\relax
\EndOfBibitem
\bibitem[Bolvin and Autschbach(2014)Bolvin, and Autschbach]{Bolvin2014}
Bolvin,~H.; Autschbach,~J. In \emph{Handbook of Relativistic Quantum
  Chemistry}; Liu,~W., Ed.; Springer Berlin Heidelberg: Berlin, Heidelberg,
  2014; pp 1--39\relax
\mciteBstWouldAddEndPuncttrue
\mciteSetBstMidEndSepPunct{\mcitedefaultmidpunct}
{\mcitedefaultendpunct}{\mcitedefaultseppunct}\relax
\EndOfBibitem
\bibitem[Sharkas \latin{et~al.}(2015)Sharkas, Pritchard, and
  Autschbach]{Sharkas2015}
Sharkas,~K.; Pritchard,~B.; Autschbach,~J. Effects from Spin-Orbit Coupling on
  Electron-Nucleus Hyperfine Coupling Calculated at the Restricted Active Space
  Level for Kramers Doublets. \emph{J. Chem. Theory Comput.} \textbf{2015},
  \emph{11}, 538--549\relax
\mciteBstWouldAddEndPuncttrue
\mciteSetBstMidEndSepPunct{\mcitedefaultmidpunct}
{\mcitedefaultendpunct}{\mcitedefaultseppunct}\relax
\EndOfBibitem
\bibitem[Autschbach \latin{et~al.}()Autschbach, Zheng, and
  Schurko]{Autschbach2010}
Autschbach,~J.; Zheng,~S.; Schurko,~R.~W. Analysis of Electric Field Gradient
  Tensors at Quadrupolar Nuclei in Common Structural Motifs. \emph{Concepts
  Magn. Reson. Part A} \emph{36A}, 84--126\relax
\mciteBstWouldAddEndPuncttrue
\mciteSetBstMidEndSepPunct{\mcitedefaultmidpunct}
{\mcitedefaultendpunct}{\mcitedefaultseppunct}\relax
\EndOfBibitem
\bibitem[Pyykkö(2008)]{Pyykko2008}
Pyykkö,~P. Year-2008 Nuclear Quadrupole Moments. \emph{Mol. Phys.}
  \textbf{2008}, \emph{106}, 1965--1974\relax
\mciteBstWouldAddEndPuncttrue
\mciteSetBstMidEndSepPunct{\mcitedefaultmidpunct}
{\mcitedefaultendpunct}{\mcitedefaultseppunct}\relax
\EndOfBibitem
\bibitem[Aquilante \latin{et~al.}(2016)Aquilante, Autschbach, Carlson,
  Chibotaru, Delcey, De~Vico, Fdez.~Galv{\'{a}}n, Ferr{\'{e}}, Frutos,
  Gagliardi, Garavelli, Giussani, Hoyer, Li~Manni, Lischka, Ma, Malmqvist,
  M{\"{u}}ller, Nenov, Olivucci, Pedersen, Peng, Plasser, Pritchard, Reiher,
  Rivalta, Schapiro, Segarra-Mart{\'{i}}, Stenrup, Truhlar, Ungur, Valentini,
  Vancoillie, Veryazov, Vysotskiy, Weingart, Zapata, and Lindh]{Molcas}
Aquilante,~F. \latin{et~al.}  Molcas 8: New Capabilities for
  Multiconfigurational Quantum Chemical Calculations across the Periodic Table.
  \emph{J. Comput. Chem.} \textbf{2016}, \emph{37}, 506--541\relax
\mciteBstWouldAddEndPuncttrue
\mciteSetBstMidEndSepPunct{\mcitedefaultmidpunct}
{\mcitedefaultendpunct}{\mcitedefaultseppunct}\relax
\EndOfBibitem
\bibitem[Douglas and Kroll(1974)Douglas, and Kroll]{Douglass1974}
Douglas,~M.; Kroll,~N.~M. Quantum Electrodynamical Corrections to the Fine
  Structure of Helium. \emph{Ann. Phys.} \textbf{1974}, \emph{82},
  89--155\relax
\mciteBstWouldAddEndPuncttrue
\mciteSetBstMidEndSepPunct{\mcitedefaultmidpunct}
{\mcitedefaultendpunct}{\mcitedefaultseppunct}\relax
\EndOfBibitem
\bibitem[Hess(1986)]{Hess1986}
Hess,~B.~A. Relativistic Electronic-Structure Calculations Employing a
  Two-Component No-Pair Formalism with External-Field Projection Operators.
  \emph{Phys. Rev. A} \textbf{1986}, \emph{33}, 3742--3748\relax
\mciteBstWouldAddEndPuncttrue
\mciteSetBstMidEndSepPunct{\mcitedefaultmidpunct}
{\mcitedefaultendpunct}{\mcitedefaultseppunct}\relax
\EndOfBibitem
\bibitem[Widmark \latin{et~al.}(1990)Widmark, Malmqvist, and Roos]{Widmark1990}
Widmark,~P.-O.; Malmqvist,~P.-{\AA}.; Roos,~B.~O. Density Matrix Averaged
  Atomic Natural Orbital (ANO) Basis Sets for Correlated Molecular Wave
  Functions. \emph{Theor. Chem. Acc.} \textbf{1990}, \emph{77}, 291--306\relax
\mciteBstWouldAddEndPuncttrue
\mciteSetBstMidEndSepPunct{\mcitedefaultmidpunct}
{\mcitedefaultendpunct}{\mcitedefaultseppunct}\relax
\EndOfBibitem
\bibitem[Roos \latin{et~al.}(2004)Roos, Lindh, Malmqvist, Veryazov, and
  Widmark]{Roos2004}
Roos,~B.~O.; Lindh,~R.; Malmqvist,~P.-{\AA}.; Veryazov,~V.; Widmark,~P.-O. Main
  Group Atoms and Dimers Studied with a New Relativistic ANO Basis Set.
  \emph{J. Phys. Chem. A} \textbf{2004}, \emph{108}, 2851--2858\relax
\mciteBstWouldAddEndPuncttrue
\mciteSetBstMidEndSepPunct{\mcitedefaultmidpunct}
{\mcitedefaultendpunct}{\mcitedefaultseppunct}\relax
\EndOfBibitem
\bibitem[Roos \latin{et~al.}(1980)Roos, Taylor, and Siegbahn]{Roos1980}
Roos,~B.~O.; Taylor,~P.~R.; Siegbahn,~P. E.~M. A Complete Active Space SCF
  Method (CASSCF) Using a Density Matrix Formulated Super-CI Approach.
  \emph{Chem. Phys.} \textbf{1980}, \emph{48}, 157--173\relax
\mciteBstWouldAddEndPuncttrue
\mciteSetBstMidEndSepPunct{\mcitedefaultmidpunct}
{\mcitedefaultendpunct}{\mcitedefaultseppunct}\relax
\EndOfBibitem
\bibitem[Siegbahn \latin{et~al.}(1981)Siegbahn, Alml{\"{o}}f, Heiberg, and
  Roos]{Siegbahn1981}
Siegbahn,~P. E.~M.; Alml{\"{o}}f,~J.; Heiberg,~A.; Roos,~B.~O. The Complete
  Active Space SCF (CASSCF) Method in a Newton-Raphson Formulation with
  Application to the HNO Molecule. \emph{J. Chem. Phys.} \textbf{1981},
  \emph{74}, 2384--2396\relax
\mciteBstWouldAddEndPuncttrue
\mciteSetBstMidEndSepPunct{\mcitedefaultmidpunct}
{\mcitedefaultendpunct}{\mcitedefaultseppunct}\relax
\EndOfBibitem
\bibitem[Hess \latin{et~al.}(1996)Hess, Marian, Wahlgren, and Gropen]{Hess1996}
Hess,~B.~A.; Marian,~C.~M.; Wahlgren,~U.; Gropen,~O. A Mean-Field Spin-Orbit
  Method Applicable to Correlated Wavefunctions. \emph{Chem. Phys. Lett.}
  \textbf{1996}, \emph{251}, 365 -- 371\relax
\mciteBstWouldAddEndPuncttrue
\mciteSetBstMidEndSepPunct{\mcitedefaultmidpunct}
{\mcitedefaultendpunct}{\mcitedefaultseppunct}\relax
\EndOfBibitem
\bibitem[Malmqvist \latin{et~al.}(2002)Malmqvist, Roos, and
  Schimmelpfennig]{rassi}
Malmqvist,~P.-{\AA}.; Roos,~B.~O.; Schimmelpfennig,~B. The Restricted Active
  Space (RAS) State Interaction Approach with Spin-Orbit Coupling. \emph{Chem.
  Phys. Lett.} \textbf{2002}, \emph{357}, 230--240\relax
\mciteBstWouldAddEndPuncttrue
\mciteSetBstMidEndSepPunct{\mcitedefaultmidpunct}
{\mcitedefaultendpunct}{\mcitedefaultseppunct}\relax
\EndOfBibitem
\bibitem[Lan \latin{et~al.}(2014)Lan, Kurashige, and Yanai]{Lan2014}
Lan,~T.~N.; Kurashige,~Y.; Yanai,~T. Toward Reliable Prediction of Hyperfine
  Coupling Constants Using Ab Initio Density Matrix Renormalization Group
  Method: Diatomic 2Σ and Vinyl Radicals as Test Cases. \emph{J. Chem. Theory
  Comput.} \textbf{2014}, \emph{10}, 1953--1967\relax
\mciteBstWouldAddEndPuncttrue
\mciteSetBstMidEndSepPunct{\mcitedefaultmidpunct}
{\mcitedefaultendpunct}{\mcitedefaultseppunct}\relax
\EndOfBibitem
\bibitem[Griffith(1963)]{Griffith1963}
Griffith,~J.~S. Spin Hamiltonian for Even-Electron Systems Having Even
  Multiplicity. \emph{Phys. Rev.} \textbf{1963}, \emph{132}, 316--319\relax
\mciteBstWouldAddEndPuncttrue
\mciteSetBstMidEndSepPunct{\mcitedefaultmidpunct}
{\mcitedefaultendpunct}{\mcitedefaultseppunct}\relax
\EndOfBibitem
\bibitem[Wernsdorfer \latin{et~al.}(2000)Wernsdorfer, Sessoli, Caneschi,
  Gatteschi, Cornia, and Mailly]{Wernsdorfer2000}
Wernsdorfer,~W.; Sessoli,~R.; Caneschi,~A.; Gatteschi,~D.; Cornia,~A.;
  Mailly,~D. Landau-Zener Method to Study Quantum Phase Interference of Fe$_8$
  Molecular Nanomagnets (Invited). \emph{J. Appl. Phys.} \textbf{2000},
  \emph{87}, 5481--5486\relax
\mciteBstWouldAddEndPuncttrue
\mciteSetBstMidEndSepPunct{\mcitedefaultmidpunct}
{\mcitedefaultendpunct}{\mcitedefaultseppunct}\relax
\EndOfBibitem
\bibitem[Taran \latin{et~al.}(2019)Taran, Bonet, and Wernsdorfer]{Taran2019}
Taran,~G.; Bonet,~E.; Wernsdorfer,~W. The Role of the Quadrupolar Interaction
  in the Tunneling Dynamics of Lanthanide Molecular Magnets. \emph{J. Appl.
  Phys.} \textbf{2019}, \emph{125}, 142903\relax
\mciteBstWouldAddEndPuncttrue
\mciteSetBstMidEndSepPunct{\mcitedefaultmidpunct}
{\mcitedefaultendpunct}{\mcitedefaultseppunct}\relax
\EndOfBibitem
\end{mcitethebibliography}

\pagebreak

\section*{Synopsis}
Hyperfine interactions for $^{159}$Tb nucleus in TbPc$_2$ single-molecule magnets are investigated from first principles using multireference calculations. Strong nuclear quadrupole and magnetic hyperfine coupling are found with the latter being dominated by the paramagnetic spin-orbital mechanism.  We construct an \emph{ab initio} pseudo-spin Hamiltonian and obtain the electronic-nuclear spectrum that is in excellent agreement with experiment. The Zeeman diagram is calculated and the magnetization dynamics is discussed. The effects of molecular distortions are studied.
\end{document}